\documentclass[aps,amsmath,superscriptaddress,amssymb,twocolumn]{revtex4-2}  
\usepackage{color}
\usepackage{enumitem}
\usepackage{graphicx,subfigure}
\usepackage[normalem]{ulem}
\usepackage{lineno}
\usepackage{booktabs}
   \usepackage{array}
    \usepackage{multirow}
    \usepackage{hhline}
    \usepackage{makecell}
     \usepackage[symbol]{footmisc}

\newif\ifHighlitedChanges
\def\ifHighlitedChanges{\iftrue}
\ifHighlitedChanges
   
  \def\STRIKE#1{{\color{blue}\sout{#1}}}
\else
  
  \def\STRIKE#1{\relax}
\fi

\newif\ifHighlitedChanges
\def\ifHighlitedChanges{\iftrue}
\ifHighlitedChanges

\else

\fi

\begin{document}
\title{Manifestations of the structural origin of supercooled water’s anomalies in the heterogeneous relaxation on the potential energy landscape}
\author{Arijit Mondal}\thanks{equal contribution}
\affiliation{Department of Chemistry, Indian Institute of Science Education and Research (IISER) Tirupati, Tirupati, Andhra Pradesh 517507, India}
\author{Gadha R.}\thanks{equal contribution}
\affiliation{Department of Physics, Indian Institute of Science Education and Research (IISER) Tirupati, Tirupati, Andhra Pradesh 517507, India}
\author{Rakesh S. Singh}
\email{rssingh@iisertirupati.ac.in}
\affiliation{Department of Chemistry, Indian Institute of Science Education and Research (IISER) Tirupati, Tirupati, Andhra Pradesh 517507, India}
\date{\today}
\begin{abstract}
Liquid water is well-known for its intriguing thermodynamic anomalies in the supercooled state. The phenomenological two-state models -- based on the assumption of the existence of two types of competing local states (or, structures) in liquid water -- have been extremely successful in describing water’s thermodynamic anomalies. However, the precise structural features of these competing local states in liquid water still remain elusive. Here, we have employed a geometrical order parameter free approach to unambiguously identify the two types of competing local states --- entropically and energetically-favored --- with significantly different structural and energetic features in the TIP4P/2005 liquid water. This identification is based on the heterogeneous structural relaxation of the system in the potential energy landscape (PEL) during the steepest-descent potential energy minimization. This heterogeneous relaxation is characterized using order parameters inspired by the spin-glass transition in frustrated magnetic systems. We have further established a direct relationship between the population fluctuation of the two states and the anomalous behavior of the heat capacity in supercooled water. The composition-dependent spatial distribution of the entropically-favored local states shows an interesting crossover from a spanning network-like single cluster to the spatially delocalized clusters in the close vicinity of the Widom line. Additionally, this study establishes a direct relationship between the topographic features of the PEL and the water's thermodynamic anomalies in the supercooled state and provides alternate markers (in addition to the locus of maxima of thermodynamic response functions) for the Widom line in the phase plane. 
\end{abstract}

\maketitle

\section{Introduction}\label{intro}
Liquid water is well-known for its anomalous thermodynamic behavior in both
the normal and the supercooled states~\cite{pablo_book, bagchi_book, pablo_rev_2003, angell_1973, angell_1982, tombari_anomaly_1999,speedy_1976, kanno_angell_1979}. The anomalous thermodynamic behavior of the supercooled water includes the sharp increase of the thermodynamic response functions (such as, isobaric heat capacity $C_P$, isothermal compressibility $\kappa_T$) on isobaric cooling. Over the last five decades, several theoretical scenarios have been proposed to interpret the anomalous behavior of supercooled water~\cite{pablo_rev_2003, speedy_1982, poole_1992, taxeria_1980, bagchi_2011}. One popular interpretation posits the existence of a hypothetical first-order liquid-liquid  transition (LLT) between the two metastable liquid phases -- high-density liquid (HDL) and low-density liquid (LDL)~\cite{poole_1992}. This proposed LLT line terminates at a liquid-liquid critical point (LLCP) in the supercooled region. As at the critical point, the thermodynamic response functions diverge, the observed anomalous behavior (divergence-like sharp increase) of the thermodynamic response functions in the supercooled water is attributed to this hypothetical LLCP and the associated Widom line, defined as the locus of maximum correlation length. Many recent computer simulation studies on atomistic water models suggest the existence of the metastable LLT~\cite{palmer_chem_rev_2018, poole_1992, pablo_nature_2014, pablo_science, abascal_2010, poole_2013, skinner_2016, jeremy_scattering_2018, pablo_wail_2022}. In a recent seminal study, Gartner \textit{et al.}~\cite{pablo_glass_2021} reported the manifestations of the LLCP in the long-range structure of the TIP4P/2005 water glass at pressures close to the critical pressure. Direct experimental verification of the LLT, however, is extremely challenging due to rapid ice crystallization, although recent state-of-the-art experiments do seem to support the existence of an LLCP in supercooled water~\cite{kim2020experimental, pathak2021enhancement}. 

Recently, two-state (or, two-structure) models have been employed extensively to understand the water's anomalies and provide a molecular basis of the LLT in supercooled water~\cite{lars_2015, tanaka_2000, soper_2000, lars_2009, anisimov_nat, anisimov_st2, anisimov_tip4p, patey_2012, patey_2013, lars_rev, tanaka_natcomm, hamm_2016, bagchi_2018, uralcan2019pattern, caupin2019thermodynamics, fausto_2019, shi2020anomalies, caupin2021minimal}. In a two-state model, the liquid water is considered as structurally heterogeneous that can be broadly classified as a mixture of two types of interconvertible local states -- energetically-favored LDL-like and entropically-favored HDL-like. The phenomenological equations of state based on the two-state models (usually referred to as two-state/structure equation of state, or TSEoS) have been extremely successful in describing the anomalous thermodynamic behavior of the real water and the water models~\cite{anisimov_nat, anisimov_st2, anisimov_tip4p, uralcan2019pattern, caupin2019thermodynamics, pablo_wail_2022} including an \textit{ab initio} deep neural network model of water~\cite{pablo_pnas_2020}. The phenomenological TSEoS-based approaches, however, do not provide molecular level insights into the detailed structural features of the two types of local states, and in turn, into the microscopic structural origin of the observed anomalous thermodynamic behavior of supercooled water.  

To understand the precise structural origin of the anomalous thermodynamic behavior of the supercooled water within the framework of a two-state model, first, we need to unambiguously identify the two types of competing local states in liquid water. In computer simulation studies, this unambiguous identification requires an order parameter that would show a pronounced bimodal distribution in liquid water, especially below the freezing temperature. Over the years, several order parameters have been proposed~\cite{chau_hardwick_1998, lsi, pablo_nature_2001, poole_d5, tanaka_natcomm, hamm_2016, v4, russo_sciortino_2021, tanaka2019revealing} to characterize the local structures of liquid water, and among these the local structure index (LSI, usually denoted by $I$) and the local translational order parameter ($\zeta$) have been used extensively to quantify the two types of local states (or, structures) in liquid water~\cite{gallo_rev_2021, lsi_spce, lars_2011, lsi_car, anisimov_tip4p, lars_lsi_2019, tanaka_natcomm, shi_tanaka_2018, shi_russo_tanaka_2018, shi_russo_tanaka_pnas_2018,gustavo_2019, shi2020anomalies}. Both the LSI and the $\zeta$ index are sensitive to the translational order up to the second coordination shell of the central molecule and provide important insights into the nature of locally-favored structures in liquid water. 

The LSI shows a clear bimodal distribution for energy minimized configurations or inherent structures (ISs), suggesting two well-defined local states in the ISs~\cite{lsi_spce, lars_2011}. However, the LSI distributions calculated for the thermally equilibrated (TE) configurations do not show bimodal features~\cite{lsi, lars_lsi_2019} (show weak bimodality at deeply supercooled conditions). The $\zeta$ index displays a bimodal distribution only for the TIP5P water~\cite{tip5p}, and a unimodal distribution for the TIP4P/2005 water (model system studied in this work)~\cite{tanaka_natcomm, shi_tanaka_2018, shi_russo_tanaka_2018, gustavo_2019}. Thus, although the LSI distinguishes clearly the two types of local states in the ISs, both of these parameters (LSI and $\zeta$) fail to unambiguously identify the two types of local states in the TE configurations. Therefore, the precise microscopic structural origin of the anomalous thermodynamic behavior of supercooled water still remains elusive. The two-state models of water are also supported by experiments~\cite{huang_pnas_2009, torre_natcomm, nilsson_2015, pettersson2022local}, however, these experiments again do not provide detailed information about the local structural features of the two states in liquid water. 

Here, using molecular dynamics simulations of the realistic TIP4P/2005 water~\cite{tip4p}, we present an alternate -- predetermined structural (or, geometrical) order parameter free -- approach to unambiguously characterize the two types of competing local states in liquid water. The two types of local states are assigned on the basis of the heterogeneous structural relaxation of the system in the potential energy landscape (PEL) during the steepest-descent energy minimization~\cite{berthier_2022}. Additionally, we have established a direct relationship between the population fluctuation of these competing local states and the anomalous temperature ($T$) dependence of the heat capacity of the system on isobaric cooling and also unravelled the signatures of this anomalous behavior encoded in the PEL of the system. 

The rest of this paper is organized as follows. Section~\ref{method} details
the computational protocol followed for simulations of the TIP4P/2005 water. In Section~\ref{sec:1} we discuss the characterization of the two types of local states in liquid water based on the local structural differences between the TE configuration and the IS of the same configuration. The interplay between the local structural fluctuations and the (anomalous) thermodynamic behavior of the TIP4P/2005 water is discussed in Section~\ref{sec:2}. In Section~\ref{sec:3}, the order parameters to characterize distinct locally relaxing regions during the energy minimization are discussed. A detailed analysis of the local energetics and structural features of the two states is presented in Sections~\ref{sec:4} \&~\ref{sec:5}, respectively. In Section~\ref{sec:6}, we have discussed the composition-dependent spatial distribution of the two states, and in Section~\ref{sec:7}, we present the phase diagram of TIP4P/2005 water summarizing the additional markers for the Widom line. The major conclusions from this work are summarized in Section~\ref{conclusions}.

\begin{figure*}
    \centering
     \begin{subfigure}
         \centering
          \includegraphics[width=0.49\linewidth]{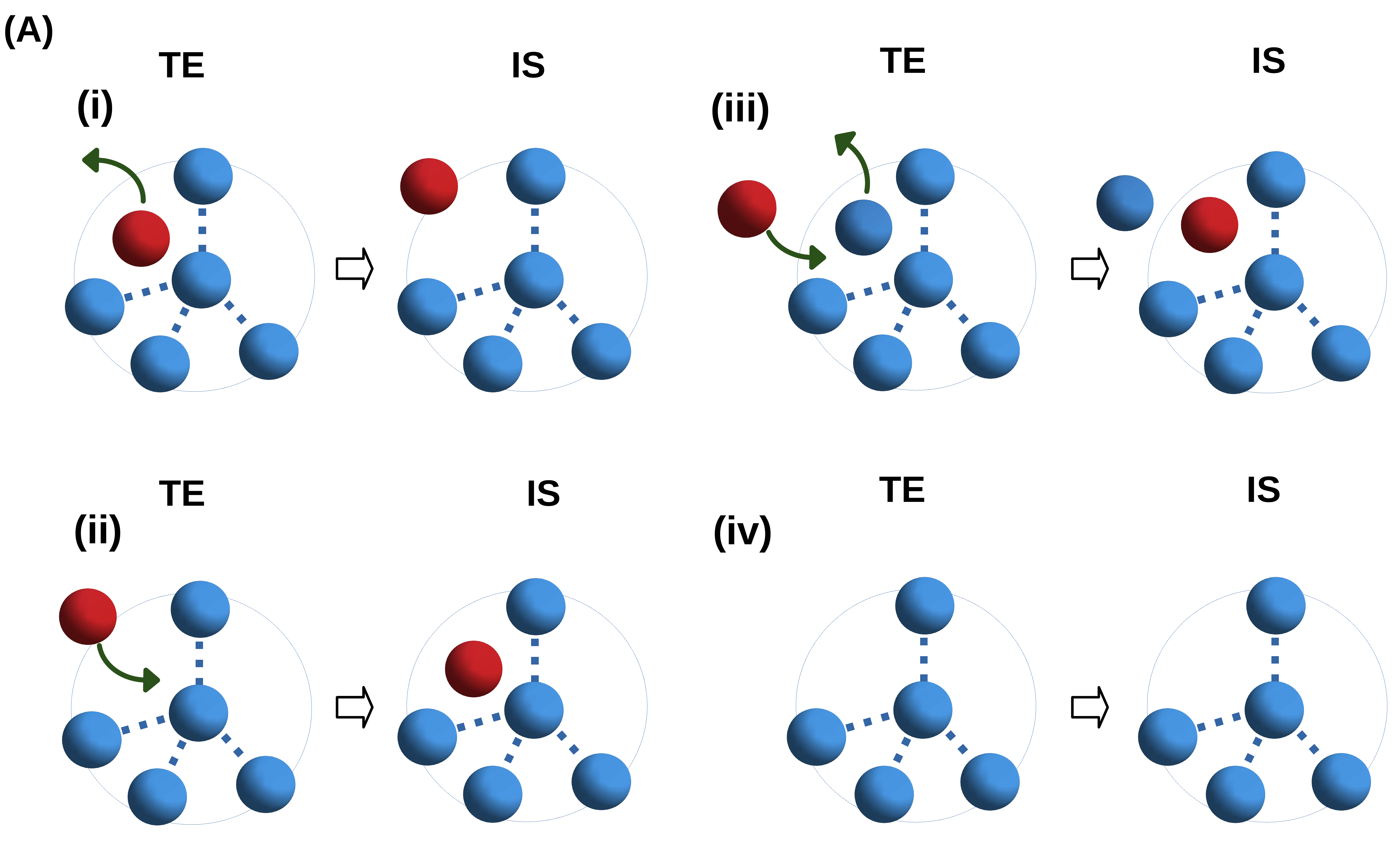}
     \end{subfigure}
     \begin{subfigure}
         \centering
           \includegraphics[width=0.49\linewidth]{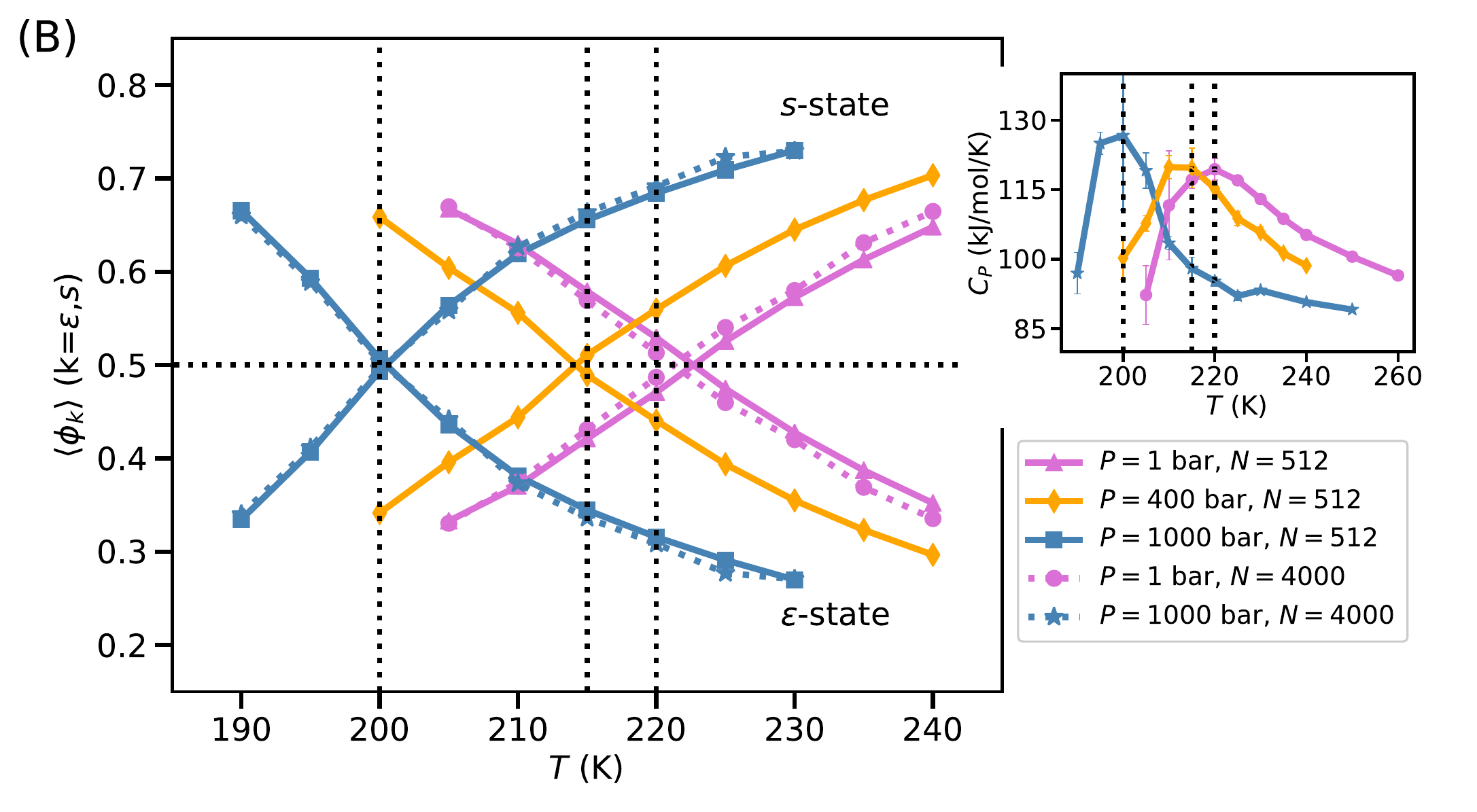}
     \end{subfigure}
    \caption{(A) A schematic representation of the different types of possible local structural rearrangements around a central water molecule during the energy minimization of the system is shown (for the sake of simplicity, only the positions of the oxygen atoms are used to specify the local structures). (i) A neighboring oxygen atom of the central water molecule in the thermally equilibrated (TE) configuration is no longer a neighbor of the same molecule in the inherent structure (IS). (ii) An oxygen atom that was not a neighbor of the central water molecule in the TE configuration is a neighbor in the IS. (iii) The central water molecule exchanges its one or more neighbors with the surrounding keeping the total number of neighboring water molecules in the IS exactly the same as in the TE configuration. (iv) The neighbor lists of the central water molecule in the TE configuration and the IS are exactly the same. (B) The $T$-dependent fraction of water molecules in the $s$- and  $\varepsilon$-states at pressures $1$ bar, $400$ bar, and $1000$ bar in the TE configurations is shown. Note the $1:1$ fraction of the $s$-state and the $\varepsilon$-state molecules at the temperature of the maximum heat capacity, $T_{C_P^{\rm {max}}}$ ($\approx 220$ K, $212$ K, and $200$ K for isobars $1$ bar, $400$ bar, and $1000$ bar, respectively; see the inset figure).} 
    \label{defect_diagram}
\end{figure*}

\section{Simulation protocol} \label{method}
Molecular dynamics (MD) simulations of TIP4P/2005 water~\cite{tip4p} were performed with GROMACS 4.6.5 ~\cite{gromacs} for $N = 512$ and $4000$ molecule systems in the isothermal isobaric ($NPT$) ensemble at pressures $1$ bar, $400$ bar, and $1000$ bar; and temperatures ranging from $260$ K to $190$ K. The short-range Lennard-Jones (LJ) part of the potential was truncated at $0.95$ nm, and long-range corrections were applied to the short-range LJ interaction for both energy and pressure. The electrostatic interaction was truncated at $0.95$ nm, and the particle mesh Ewald was used to compute the long-range contributions to the electrostatics. In all cases, periodic boundary conditions were applied, and a time step of $2$ fs was used to propagate the trajectories. The linear constraint solver (LINCS)~\cite{lincs} algorithm was used to handle TIP4P/2005’s rigid body constraints. To maintain a constant temperature, we used N$\ddot{\rm o}$se-Hoover thermostat~\cite{nose, hoover} with $0.2$ ps relaxation time. Constant pressure was maintained using Parrinello-Rahman barostat~\cite{pr_barostat} with $2$ ps relaxation time. To ensure equilibration of the system at thermodynamic conditions studied in this work, we performed simulations varying between $100$ ns and 5 $\mu$s ($> 200 \tau_s$ at all the temperatures studied in this work. $\tau_s$ is the characteristic structural relaxation time of the system defined as the time at which the self intermediate scattering function, $F_s(k^{\ast},t) = 1/e$, $k^{\ast}$ is the wavenumber corresponding to the first peak of structure factor). 

\section{Results and Discussion}
\subsection{\label{sec:1}Two types of local states in liquid water}
Two distinct types of local states in a TE liquid water configuration are characterized by mapping the TE configuration onto its IS (local minimum of the PEL)~\cite{berthier_2022}. The IS of a TE configuration is obtained by energy minimization of the TE configuration using the steepest-descent method~\cite{weber_1982, stillinger_book}. During the energy minimization, the system evolves on a multidimensional potential energy surface, and different water molecules undergo different degrees of local structural rearrangements. Figure~\ref{defect_diagram}A shows a schematic representation of the different types of possible local structural rearrangements a water molecule can undergo during the energy minimization of the system. We define the local structure of a water molecule by its neighbor list (list of nearest neighbor molecules). We have used oxygen-oxygen radial cut-off distance ($r_{OO}$) of $3.7~\AA$ to calculate the neighbor list of a central water molecule. This choice of cut-off distance is motivated by the recent findings that one must include the structural information beyond the first shell ($3.5~\AA$) to get the signatures of the two types of local states in liquid water~\cite{lsi_spce,lars_2011,tanaka_natcomm} (sensitivity of the results on the choice of $r_{OO}$ is discussed in Section~\ref{subsec:level1} of the Supplementary Materials). There are three different ways a central water molecule can change its neighbor list during the energy minimization:  (a) it loses some of its neighboring molecules without gaining any new neighbor(s) (Fig.~\ref{defect_diagram}A(i)), (b) it gains new neighbors without losing any neighbors (Fig.~\ref{defect_diagram}A(ii)), and (c) it exchanges its one or more neighbors with the surrounding (Fig.~\ref{defect_diagram}A(iii)). 

\begin{figure} [h]
    \centering
     \includegraphics[width=0.85\linewidth]{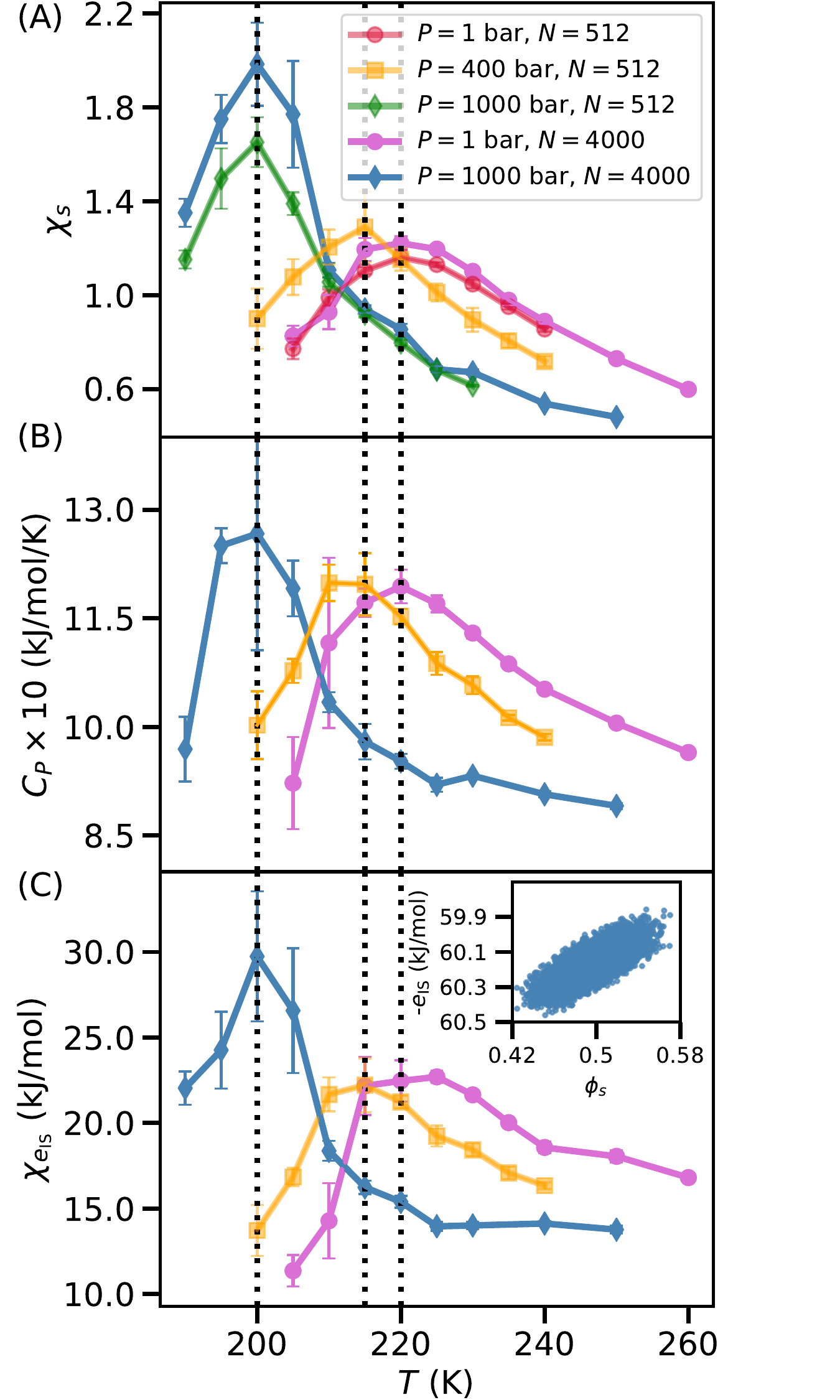}
    \caption{The $s$-state population fluctuation, $\chi_s$ (A), the heat capacity at constant pressure, $C_P$ (B), and the fluctuation of the inherent structure energy per molecule ($e_{\rm IS}$), $\chi_{e_{\rm IS}}$ (C) on isobaric cooling at pressures $1$ bar, $400$ bar, and $1000$ bar are shown. We note that the temperature of the maximum of the $s$-state population fluctuation coincides nearly perfectly with the $C_P$ maximum and the $\chi_{e_{\rm IS}}$ maximum for all the three isobars studied in this work. In the inset, we have shown the correlation between the $e_{\rm IS}$ and the $s$-state population $\phi_s$ at pressure $1000$ bar and temperature $200$ K ($T_{C_P^{\rm {max}}}$ for $1000$ bar isobar).}
    \label{figure2}
\end{figure}

To quantify the local structural evolution during the energy minimization, we define a parameter $\phi$ for each water molecule, such that $\phi_{i}=0$ if molecule $i$ neither loses nor gains any neighbor(s) during the relaxation in the PEL (\textit{i.e.,} the same water molecules are neighbors in both the TE configuration and the IS), and $\phi_{i} = 1$ otherwise. This binary classification enables us to unambiguously distinguish the two types of local states in the TE configurations. This approach has recently been used to study the role of localized defects on the relaxation dynamics in the PEL of glass-forming liquids~\cite{berthier_2022}. A water molecule with $\phi = 1$ is assigned as an entropically-favored local state (or, $s$-state) in the TE configuration as it undergoes significant local structural change during the energy minimization (see Figs.~\ref{defect_diagram}A(i-iii)); and a water molecule with $\phi = 0$ is assigned as an energetically-favored local state (or, $\varepsilon$-state) in the TE configuration as it undergoes negligible local structural change (no change in the neighbor list) on energy minimization (see Fig.~\ref{defect_diagram}A(iv)). The majority of the $s$-state molecules (ca. $70 \%$ at the temperature where $C_P$ shows a maximum on isobaric cooling, $T_{C_P^{\rm {max}}}$; see the inset of Fig.~\ref{defect_diagram}B) lose a nearest neighbor water molecule and become tetrahedrally coordinated after the energy minimization (see Fig.~\ref{fig2xs} in the Supplementary Materials).  

Unlike binary mixtures, the fractions of the $s$- and $\varepsilon$-states in liquid water are controlled by the thermodynamic equilibrium between these two states, and hence, depend strongly on temperature. In Fig.~\ref{defect_diagram}B, we show the ensemble average fraction of the $s$-state ($\left< \phi_s\right>$) and the $\varepsilon$-state ($\left<\phi_\varepsilon\right>$) water molecules in the TE configurations as a function of temperature at pressures $1$ bar, $400$ bar, and $1000$ bar. $\left< \phi_s\right>$ is defined as $\left< \phi_s \right> = \left< \frac{1}{N} \sum_i \phi_i \right>$, and $\left< \phi_\varepsilon \right> = 1- \left< \phi_s \right>$; here $\left<..\right>$ denotes the ensemble average. It is evident from the figure that the $s$-state population decreases and the $\varepsilon$-state population increases monotonically with the decrease in temperature. We also note that the temperature corresponding to the $1:1$ fraction of the $s$- and the $\varepsilon$-state water molecules coincides nearly perfectly with the $T_{C_P^{\rm {max}}}$ ($\approx 220$ K, $212$ K and $200$ K for $1$ bar, $400$ bar, and $1000$ bar pressure, respectively; see the inset of Fig.~\ref{defect_diagram}B). This $1:1$ population ratio suggests maximal local structural ($s$- or $\varepsilon$-state) fluctuations in the close vicinity of the $T_{C_P^{\rm {max}}}$, or the Widom line (which is commonly defined as the locus of maximum of $C_P$ or $\kappa_T$ on isobaric cooling).  A similar $1:1$ population ratio of the two states identified on the basis of the LSI and the $\zeta$ index in liquid water has previously been used to locate the Widom line (more precisely, the locus of $\kappa_T$ maximum line) in the $P-T$ plane. However, the LSI provides $1:1$ population ratio only in the ISs~\cite{lsi_spce, lars_2011}, not in the TE configurations. The $\zeta$ index distributions do not display bimodality for the TIP4P/2005 water, and the population of the two states is obtained by fitting the unimodal $\zeta$ distribution with two Gaussian functions and estimating the areas of those two Gaussian functions~\cite{tanaka_natcomm}. Thus, the relative population of the two states predicted on the basis of the $\zeta$ index is expected to be delicately sensitive to the choice of the fitting functions. 
\subsection{\label{sec:2}$\boldsymbol s$-state population fluctuations and the heat capacity anomaly in supercooled water}
The heat capacity $C_P$ is related with the system's entropy ($S$) fluctuation as $C_P = \left(\left<S^{2}\right> - \left<S\right>^2\right) / k_{\rm B}$, or the enthalpy ($H$) fluctuation as $\left(\left<H^{2}\right> - \left<H\right>^{2}\right) / {k_{\rm B} T^2}$, where $k_{\rm B}$ is Boltzmann's constant. As the enthalpy and entropy of a system depend strongly on the nature of the locally-favored structures in the system, it is expected that the anomalous behavior of the heat capacity (or, thermodynamic response functions, in general) can be directly connected with the local structural fluctuations. To establish this relationship, in Figs.~\ref{figure2}A and~\ref{figure2}B, we report the population fluctuation of water molecules in the $s$-state ($\chi_s$) along with the heat capacity $C_P$ as a function of temperature for three different isobars -- $1$ bar, $400$ bar, and $1000$ bar. $\chi_{s}$ is defined in terms of the population of the $s$-state water $\phi_s$ as $N\left(\left<\phi_s^{2}\right> - \left<\phi_s\right>^{2}\right)$, and $C_P$ is calculated from the enthalpy fluctuation of the system. 

We observe a striking similarity in the $T$-dependent behavior of $C_P$ and $\chi_{s}$ for all the three isobars (see Figs.~\ref{figure2}A and 2B). The $s$-state population susceptibility $\chi_{s}$ shows a maximum exactly at the temperature where $C_P$ shows a maximum, $T_{C_P^{\rm {max}}}$, on isobaric cooling ($\approx 220$ K, $212$ K, and $200$ K for isobars $1$ bar, $400$ bar, and $1000$ bar, respectively). Also, like $C_P$,  $\chi_{s}$ increases more sharply on isobaric cooling at higher pressures, suggesting a possible divergence (in the thermodynamic limit) at the LLCP (ca. $182$ K and $1.7$ kbar for TIP4P/2005 water~\cite{ anisimov_tip4p}). These observed striking similarities in the $T$-dependent behavior of $\chi_s$ and $C_P$ suggest that the enhanced population fluctuations of the entropically-favored $s$-state molecules on isobaric cooling give rise to the enhanced entropy fluctuations of the system, which in turn give rise to the heat capacity anomaly. We also checked the system size dependence of the observed similarities between the $T$-dependent behavior of $C_P$ and $\chi_{s}$ by performing simulations on a larger system size consisting of $N = 4000$ water molecules at pressures $1$ bar and $1000$ bar, and found that the results hold true (see Fig.~\ref{figure2}A). 

We further explored the signatures of local structural, or the $s$-state population fluctuations and the heat capacity anomaly encoded in the PEL of the system. We calculated the IS energy per molecule ($e_{\rm IS}$) fluctuation ($\chi_{e_{\rm IS}}$), defined as $\chi_{e_{\rm IS}} = \left<e_{\rm IS}^{2}\right> - \left<e_{\rm IS}\right>^2$ on isobaric cooling (Fig.~\ref{figure2}C), and compared the $T$-dependent behavior of $\chi_{e_{\rm IS}}$ with $\chi_s$ and $C_P$. We again note a striking similarity between the $T$-dependent behavior of $\chi_{e_{\rm IS}}$, $\chi_s$ and $C_P$. The position of the $\chi_{e_{\rm IS}}$ maximum coincides nearly perfectly with the maximum of $\chi_s$ and $C_P$ for all the three isobars studied in this work. Also, like $\chi_{s}$ and $C_P$, $\chi_{e_{\rm IS}}$ increases more sharply on isobaric cooling at higher pressures. These observations suggest a direct relationship between the $s$-state population in a TE configuration and the energy of the corresponding IS basin in the PEL. The TE configurations with a higher $s$-state population relax to the higher energy basins and with a lower $s$-state population relax to the lower energy basins in the PEL on energy minimization (see the inset of Fig.~\ref{figure2}C and Fig.~\ref{fig2s}A in the Supplementary Materials). That is, the behavior of the (thermal) system at higher temperatures, $T > T_{C_P^{\rm {max}}}$, with higher population of the $s$-state molecules, is strongly influenced by the higher potential energy basins in the underlying PEL, and vice-versa. Near the $T_{C_P^{\rm {max}}}$, the enhanced IS energy fluctuations suggest that both the higher and the lower energy basins in the PEL influence the system's behavior, giving rise to the enhanced $s$-state population fluctuations, and in turn, the heat capacity anomaly. Thus, the information about the anomalous temperature dependence of $C_P$ is encoded in the change in the topography of the accessible regions of the underlying PEL on isobaric cooling. Furthermore, $\chi_{e_{\rm IS}}$ can also be used as an alternate marker for the locus of the heat capacity maxima, or the Widom line, in the phase plane.  
\subsection{\label{sec:3}Order parameters to characterize the heterogeneous relaxation in the PEL}
As discussed in Section~\ref{sec:1}, during the energy minimization, different water molecules in a TE configuration undergo different degrees of local structural rearrangements, suggesting heterogeneous structural relaxation in the PEL. To further characterize this heterogeneous local relaxation in the PEL, here, we have used the translation overlap between the two replicas -- TE configuration and the corresponding IS -- as an order parameter. The translational overlap parameter ($q_t$) for molecule $i$ is defined as $q_t(i) = w \left(|\mathbf{r}^{\rm TE} (i) -  \mathbf{r}^{\rm IS} (i) |\right)$, where $\mathbf{r}^{\rm TE} (i)$ corresponds to the coordinate of the oxygen of molecule $i$ in the TE configuration, and $\mathbf{r}^{\rm IS} (i)$ corresponds to the same in the IS. $w(x)$ is a window function, which is $1$ if $x < 0.2 \sigma$ ($\sigma$ is the diameter of the water molecule estimated from the position of the first peak of the radial distribution function, $2.75~\AA$) and zero otherwise. The structures of a TE configuration and the corresponding IS are similar (translationally) if $\bar{q_t} =  \sum_{i} q_t(i) / N \approx 1$, and unrelated otherwise when molecules have moved $>0.2\sigma$ during the energy minimization. The particle average $q_t$ for the $k$-state (with $k = s$ or $\varepsilon$) molecules is given as $\bar{q_t}^k = \sum_{i \in k} q_t(i) / N_k$, where $N_k$ is the number of water molecules in state $k$.  

\begin{figure}
    \centering
     \includegraphics[width=\linewidth]{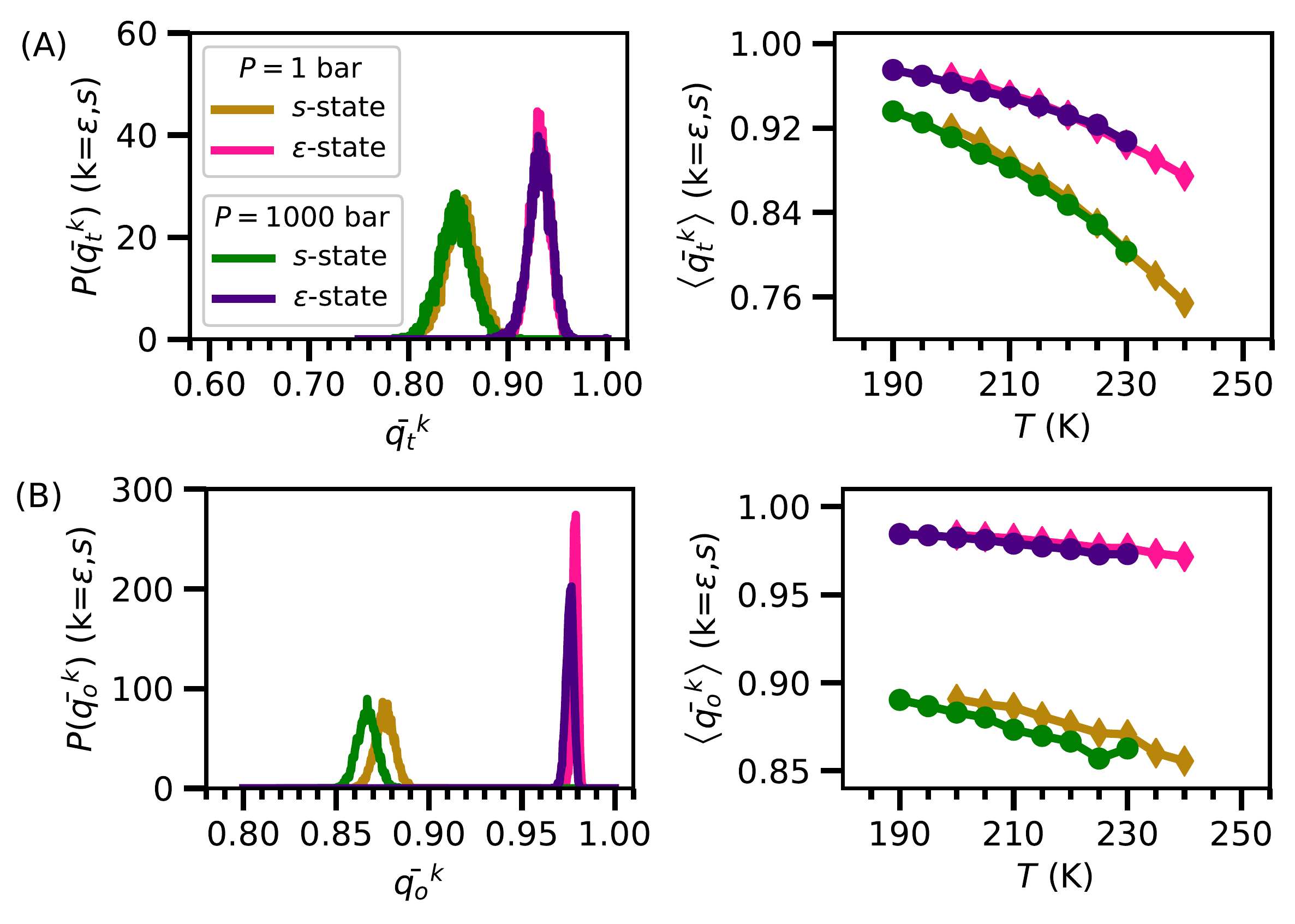}
    \caption{(A) The distribution of the particle average translational overlap parameter $\bar{q_t}$ for the $s$- and $\varepsilon$-state molecules ($P(\bar{q_t}^k)$, $k = s$ or $\varepsilon$) at the $T_{C_P^{\rm max}}$ ($220$ K and $200$ K for $1$ bar and $1000$ bar isobars, respectively); and the $T$-dependent ensemble average $\bar{q_t}$ for the $s$- and $\varepsilon$-state molecules ($\langle \bar{q_t}^k \rangle$, $k = s$ or $\varepsilon$) at pressures $1$ bar and $1000$ bar for $N = 4000$ are shown. (B) We show the same for the local orientational overlap parameter, $\bar{q_o}$. We note the complete separation of the overlap parameter (translational and orientational) distributions at the $T_{C_P^{\rm max}}$, and distinctly different values of the ensemble average overlap parameters for the $s$- and $\varepsilon$ states. These observations suggest the presence of two distinct locally relaxing environments during the steepest-descent energy minimization of the system.}
    \label{spinglass}
\end{figure}
 \begin{figure*}
    \centering
     \begin{subfigure}
         \centering
        \includegraphics[width=0.49\linewidth]{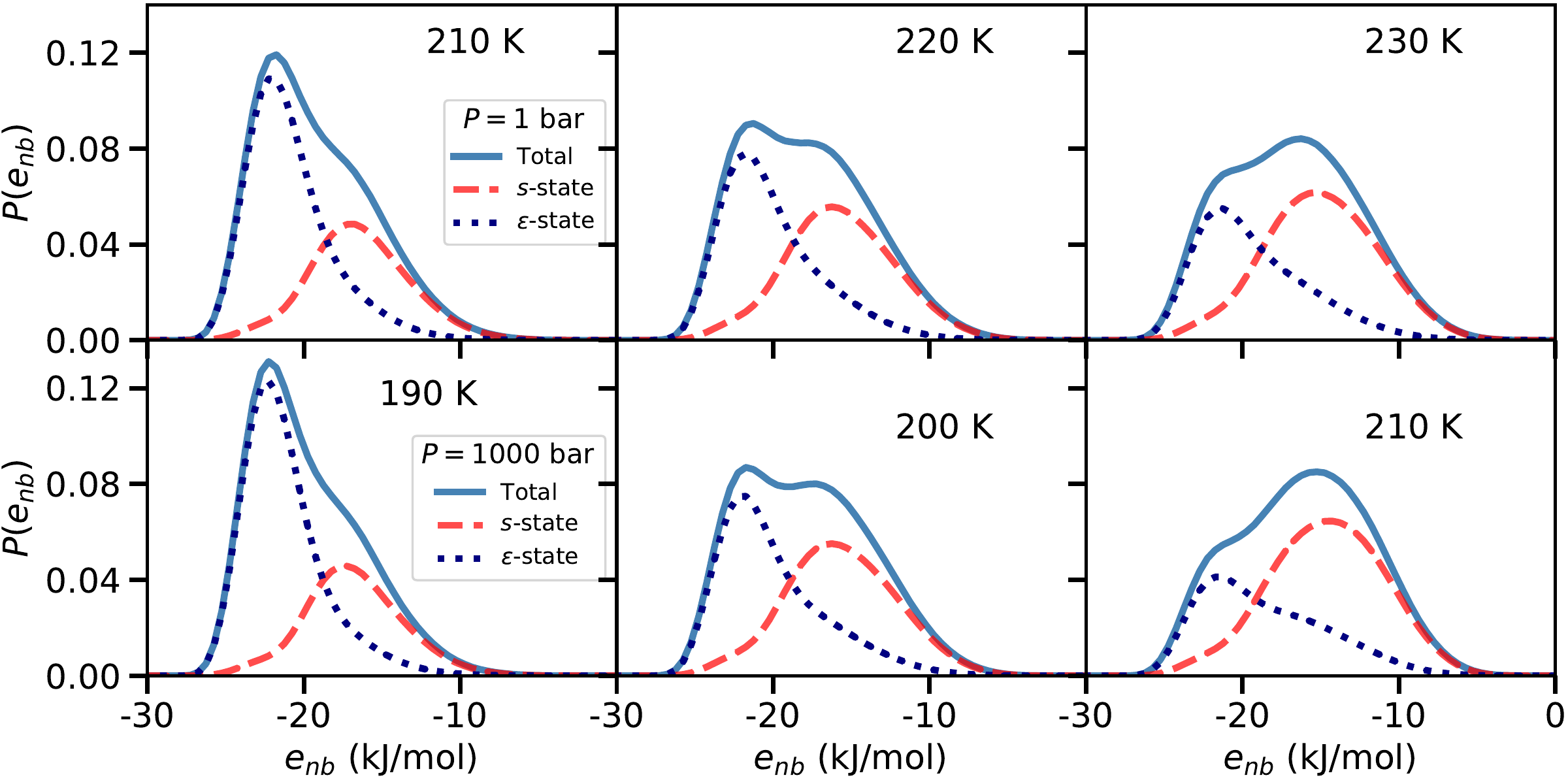}
     \end{subfigure}
     \begin{subfigure}
         \centering
         \includegraphics[width=0.49\linewidth]{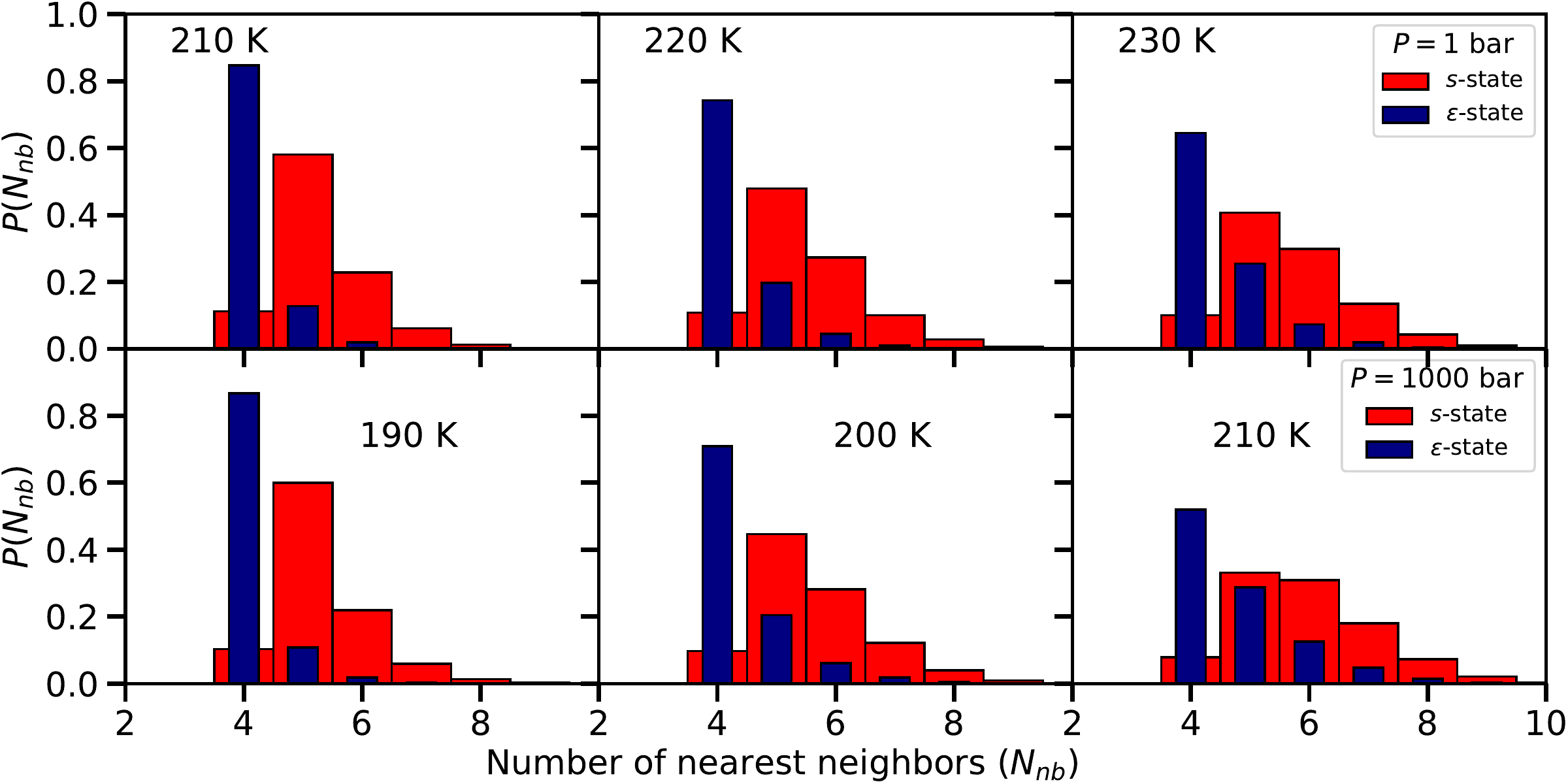}
     \end{subfigure}
    \caption{(left) The potential energy per neighbor distribution ($P(e_{nb})$) for the system and for the molecules in the $s$- and $\varepsilon$-states for isobars $1$ bar (top panel) and $1000$ bar (bottom panel) and at temperatures in the close vicinity of their respective $T_{C_P^{\rm max}}$ -- $220$ K and $200$ K. (right) The number of nearest neighbors distribution ($P(N_{nb})$) for the $s$-state and the $\varepsilon$-state molecules is shown at the same thermodynamic conditions. It is evident from the figure that the $\varepsilon$-state molecules have fewer nearest neighbors than the $s$-state, suggesting that $s$-state molecules have higher local density than the $\varepsilon$-state molecules. The system consists of $N = 4000$ water molecules.}
    \label{local_e_n}
\end{figure*}

In Fig.~\ref{spinglass}A, we show the distribution of $\bar{q_t}$ for the $s$-state and the $\varepsilon$-state molecules ($P(\bar{q_t}^k)$ with $k = s$ and $\varepsilon$, respectively) at the $T_{C_P^{\rm max}}$ for $1$ bar and $1000$ bar isobars along with the dependence of the ensemble average $\bar{q_t}$ for the $s$- and $\varepsilon$-state molecules (indicated by $\langle \bar{q_t}^s \rangle$ and $\langle \bar{q_t}^{\varepsilon} \rangle$, respectively) on temperature. We note that the $\langle \bar{q_t} \rangle$ values for the $\varepsilon$-state are higher than the $s$-state water molecules at all the temperatures. Also, the distribution of $\bar{q_t}$ for the $s$-state is shifted towards the lower $\bar{q_t}$ values and is well-separated from the $\varepsilon$-state distribution at the $T_{C_P^{\rm max}}$ (Fig.~\ref{spinglass}A, left). These observations again suggest two distinct types of locally relaxing environments -- the molecules in the $\varepsilon$-state undergo negligible or lesser translational rearrangements compared to the $s$-state molecules during the energy minimization of the system. This further suggests that the $\varepsilon$-state molecules are structurally locally rigid compared to the $s$-state molecules. 

Next, we measured the local structural (or, hydrogen-bond network) rigidity around a central water molecule in the TE configurations by introducing a local orientational overlap parameter ($q_o$) -- defined as the overlap between the net dipole moment unit vector of the water molecules participating in the local structure in the TE configuration ($\mathbf{e}_{\rm ld}^{\rm TE}$) and in the corresponding IS ($\mathbf{e}_{\rm ld}^{\rm IS}$). This definition of $q_o$ is motivated by the Edwards-Anderson spin-glass order parameter~\cite{ea}, and is mathematically defined as $q_o(i) = \mathbf{e}_{\rm ld}^{\rm TE} (i).\mathbf{e}_{\rm ld}^{\rm IS} (i)$. Here $\mathbf{e}_{\rm ld}$ represents the resultant local unit dipole moment vector of all the nearest neighbor water molecules surrounding the central water molecule $i$ (water molecules with $r_{OO} < 3.7~\AA$ from the central molecule), including the central water, and is given as $\mathbf{e}_{\rm ld} (i) = \sum_{j = 0}^{N_b(i)} \mathbf{e}_{\rm d}(j) / \sqrt{\sum_{j = 0}^{N_b(i)} \mathbf{e}_{\rm d}(j). \sum_{j = 0}^{N_b(i)} \mathbf{e}_{\rm d}(j)}$; $\mathbf{e}_{\rm d} (j)$ is dipole moment unit vector of the $j^{\rm th}$ molecule ($j=0$ stands for the central water molecule, $i$). $q_o$ is defined in such a way that it has a maximum value of $1$ for local structures undergoing no change during the relaxation in the PEL, and $ < 1$ otherwise. In Fig.~\ref{spinglass}B, we show the distribution of the particle average orientational overlap parameter, $\bar{q_o}$ for the $s$- and $\varepsilon$-state molecules (denoted as $P(\bar{q_o}^k)$ with $k = s$ and $\varepsilon$, respectively) at the $T_{C_P^{\rm max}}$ for $1$ bar and $1000$ bar isobars along with the dependence of the ensemble average orientational overlap parameter $\langle \bar{q_o}^k \rangle$ on temperature. As evident from the figure, the local structure around an $s$-state molecule undergoes significantly more orientational rearrangement compared to the $\varepsilon$-state water molecules during the energy minimization, and hence, strengthens our earlier arguments that $\varepsilon$-states are energetically-favored locally rigid (ice-like) structures, and $s$-states are entropically-favored locally flexible structures in the TE configurations. 
\begin{figure*}
    \centering
     \begin{subfigure}
         \centering
          \includegraphics[width=0.49\linewidth]{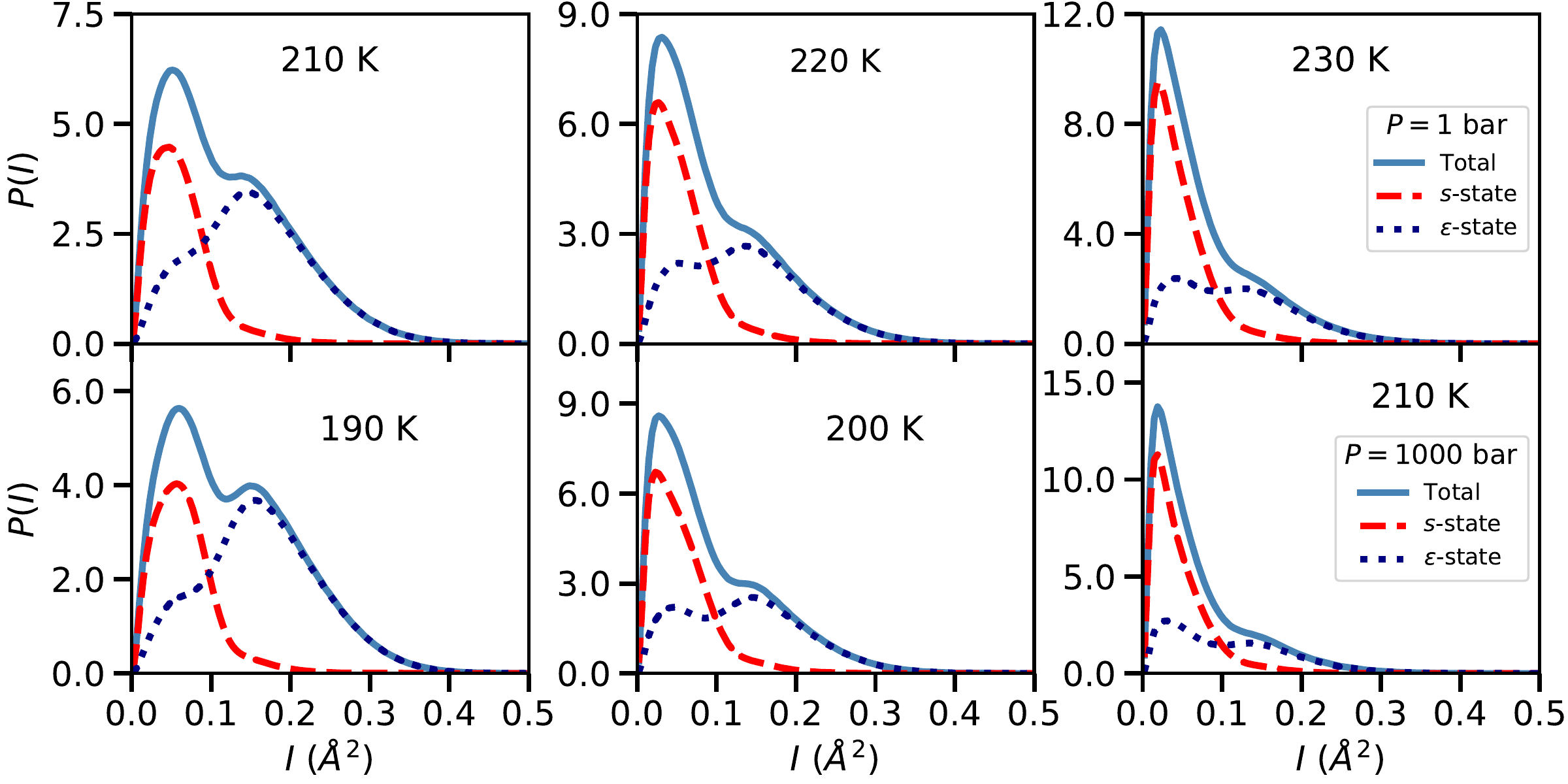}
     \end{subfigure}
     \begin{subfigure}
         \centering
           \includegraphics[width=0.49\linewidth]{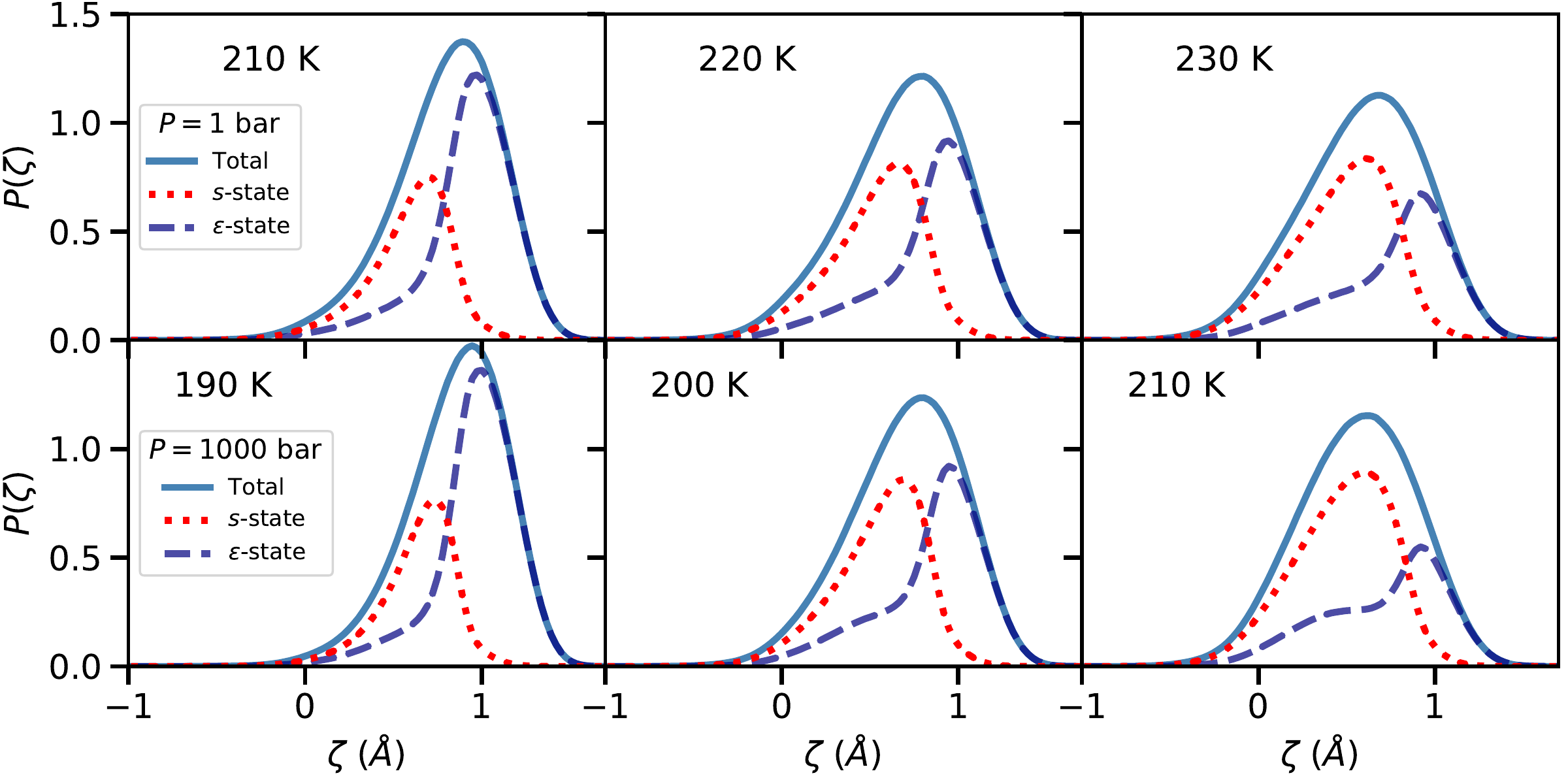}
     \end{subfigure}
    \caption{(left) The LSI ($I$) distribution and (right) the translational order parameter ($\zeta$) distribution for the system and for the molecules in the $s$- and $\varepsilon$-states for isobars $1$ bar (top panel) and $1000$ bar (bottom panel) and at temperatures in the close vicinity of their respective $T_{C_P^{\rm max}}$ -- $220$ K and $200$ K. The system consists of $N = 4000$ water molecules.} 
    \label{lsi_tanaka}
\end{figure*} 
\subsection{\label{sec:4}Energetic features and local density of the $\boldsymbol s$- and $\boldsymbol \varepsilon$-states}
To gain a deeper understanding of the energetic stability of the $s$- and $\varepsilon$-state water molecules in the TE configurations, in Fig.~\ref{local_e_n} (left), we show the local potential energy per neighbor ($e_{nb}$) distribution ($P(e_{nb})$) for the system and for the molecules in the $s$- and $\varepsilon$-states for $1$ bar and $1000$ bar isobars and at temperatures in the close vicinity of their respective $T_{C_P^{\rm max}}$. The $e_{nb}$ for molecule $i$ is defined as $e_{nb} (i) = \sum_{j = 1}^{N_{nb}} e_{ij} / N_{nb}$, where $e_{ij}$ is interaction energy between the molecule $i$ and its neighbor $j$, and $N_{nb}$ is the number of nearest neighbors of molecule $i$. The $P(e_{nb})$ exhibits bimodal characteristics in the close vicinity of the $T_{C_P^{\rm max}}$, with the peak positions coinciding with the peak position for the distribution of the $s$- and the $\varepsilon$-state molecules. This bimodality suggests a two-well nature of the local free energy surface, defined as $F_l(e_{nb}) = -k_{\rm B}T \ln \left[P(e_{nb})\right]$ (see Fig.~\ref{fig3s} in the Supplementary Materials) at the $T_{C_P^{\rm max}}$, and establishes distinct energetic features of the $s$- and $\varepsilon$-state molecules in liquid water. This bimodality further suggests that $e_{nb}$ can also be used as an order parameter to identify the two types of local states in liquid water. On lowering (increasing) temperature, the bimodality gradually disappears due to the dominance of the $\varepsilon(s)$-state molecules in the system. We also note that the local energy distribution for the $\varepsilon$-state molecules is shifted towards the lower energy, suggesting that the $\varepsilon$-state molecules are locally stabilized by stronger interactions than the $s$-state molecules. 

As stronger (hydrogen-bond) interactions usually favor low-density-like local structures in water, one would expect to observe a direct relationship between the strength of the local interactions and the local density around a central water molecule. The local density around a water molecule can be defined in terms of the number of neighbors $N_{nb}$ with $r_{OO} < 3.7~\AA$. In Fig.~\ref{local_e_n} (right), we show the number of neighbors distribution ($P(N_{nb})$) for the $s$-state and the $\varepsilon$-state molecules at different temperatures in the close vicinity of the $T_{C_P^{\rm max}}$. The number of neighbors (or, the local density) of the water molecules in the $s$-state is higher than the $\varepsilon$-state giving rise to higher bulk density for the configurations with higher $s$-state population (see Fig.~\ref{fig2s}B in the Supplementary Materials). The distribution for water molecules in the $\varepsilon$-state is peaked at $N_{nb}=4$ for all the temperatures reported -- suggesting a locally tetrahedral structure with well-separated first and second shells for the majority of the $\varepsilon$-state molecules. $P(N_{nb})$ for the $s$-state water molecules exhibits a maximum at $N_{nb}=5$, suggesting either the presence of an additional water molecule in between the first and second shells, or a distorted first shell with more than $4$  molecules (a detailed analysis of the structural features of the $\varepsilon$- and $s$-states is provided in the next section). With the increase of temperature the number of molecules with $N_{nb}=5$ increases due to an increase in the entropically-favored local structural excitations in the PEL. The above results also suggest that the $\varepsilon$-state can be considered as LDL-like and the $s$-state as HDL-like water molecules. 

\subsection{\label{sec:5}Structural features of the $\boldsymbol s$- and $\boldsymbol \varepsilon$-states} 
To probe the structural similarities between the two types of local states ($s$ and $\varepsilon$) identified in this work and the ones identified previously on the basis of the LSI~\cite{lars_2011, lars_lsi_2019_2, lars_lsi_2019} and the $\zeta$ index~\cite{tanaka_natcomm}, we calculated the LSI and $\zeta$ index distributions for both the $s$- and $\varepsilon$-state molecules. The LSI of a molecule $i$ is obtained by sorting the oxygen-oxygen nearest neighbor distances between the central molecule $i$ and its $j^{\rm th}$ nearest neighbor (denoted as $r_j$) in ascending order; $r_1 < r_2 < ... ~r_j~ ....< r_{n(i)} < 3.7~\AA < r_{n(i)+1}$. Then, the LSI is defined as~\cite{lsi}
\begin{equation}
 I(i) = \frac{1}{n(i)}\sum_{j=1}^{n(i)}\left[\Delta(j;i) - \bar\Delta(i)\right]^2,
\end{equation}
where $\Delta(j;i) = r_{j+1} - r_j$ and $\bar\Delta$ is the average of $\Delta(j;i)$ over all the nearest neighbors $j$ of molecule $i$. The LSI is a measure of inhomogeneity in between the first and second coordination shells and thus is sensitive to the translational order up to the second shell of the central molecule. The $\zeta$ index of a molecule $i$ is defined as the difference between the distance $d_{ji}$ of the closest neighbor molecule $j$ not hydrogen-bonded to molecule $i$, and the distance $d_{ki}$ of the farthest neighbor molecule $k$ hydrogen-bonded to molecule $i$, $\zeta(i) = d_{ji}-d_{ki}$~\cite{tanaka_natcomm}. Two water molecules are considered to be hydrogen-bonded only when the oxygen-oxygen distance is less than $3.5~\AA$, and the $H-O...O$ angle is less than $30^{\circ}$~\cite{luzar_chandler}. Similar to the LSI, the $\zeta$ index also probes the translational order up to the second coordination shell. 

Figure~\ref{lsi_tanaka} (left) shows the LSI distribution for the system, along with the $s$- and $\varepsilon$-state molecules, for isobars $1$ bar and $1000$ bar and at temperatures in the close vicinity of their respective $T_{C_P^{\rm max}}$. The LSI distribution for the $s$-state water molecules shows a peak at the low LSI value resembling HDL-like water molecules. However, we observe a (weak) bimodality in the distribution of the $\varepsilon$-state water molecules, especially at $T \ge T_{C_P^{\rm max}}$. This suggests that the $\varepsilon$-state water molecules have structural characteristics similar to both the LDL- and HDL-like water molecules assigned on the basis of the LSI, and thus, the two types of local states assigned on the basis of LSI in the TE configurations~\cite{lars_lsi_2019_2, lars_lsi_2019} are not the same as the two states ($s$ and $\varepsilon$) assigned in this work. In addition, we also calculated the LSI distribution for the $s$- and $\varepsilon$-state molecules in the ISs to probe the structural similarities between the two types of local states identified in this work and the ones identified on the basis of the LSI in the PEL~\cite{lars_2011}. We found a bimodal distribution for both the states (See Fig.~\ref{fig5s} in the Supplementary Materials), suggesting that the structural features of the $\varepsilon$- and $s$-state molecules again resemble only partially the LDL- and HDL-like molecules, respectively, assigned using the LSI~\cite{lars_2011} in the ISs.

Figure~\ref{lsi_tanaka} (right) shows the $\zeta$ index distribution for the system, along with the $s$-state and the $\varepsilon$-state molecules. The $s$-state molecules have lower $\zeta$ index than the $\varepsilon$-state molecules, and also, the peaks of the distributions for these two states are well separated. This suggests a well-separated first and second shells for the $\varepsilon$-state and a collapsed/distorted second shell (giving rise to the presence of the fifth neighbor in between the first and second shells) for the $s$-state molecules (the two-body oxygen-oxygen correlation functions for the $s$- and $\varepsilon$-state molecules shown in Fig.~\ref{fig4s} of the Supplementary Materials also support the aforementioned structural features of the two states). This is in qualitative agreement with the structural features of the two states of liquid water assigned on the basis of $\zeta$ index in a recent study by Russo and Tanaka~\cite{tanaka_natcomm}. However, quantitatively, the two types of local states assigned in this work are not the same as the ones predicted by Russo and Tanaka~\cite{ tanaka_natcomm} from the $\zeta$ index distribution, as they obtained a $1:1$ fraction of the two states (based on the assumption that the $\zeta$ distribution for each state follows a Gaussian distribution) near the temperature of maximum compressibility, $T_{\kappa_T^{\rm max}}$ not at the $T_{C_P^{\rm max}}$ (this work, Fig.~\ref{defect_diagram}B) on isobaric cooling. 
\subsection{\label{sec:6}Spatial distribution of the $\boldsymbol s$- and $\boldsymbol \varepsilon$-states}
After unambiguously assigning the two distinct types of local states on the basis of the heterogeneous system's relaxation in the PEL, it is desirable to probe the temperature (or, composition) dependent spatial distribution of the $\varepsilon$- and $s$-states by employing clustering analysis. An $s$-cluster of size $n$ is defined as $n$ $s$-state water molecules connected by the neighborhood (within a cut-off distance of $3.5~\AA$). To measure the spatial localization of the $s$-state molecules, we divided the average fraction of the $s$-state molecules $\left <\phi_s \right >$ in the system into two parts -- the fraction of molecules that are part of the largest cluster ($\left < \phi_s^l \right >$) and the fraction of molecules that are not part of the largest cluster, $\left < \phi_d  \right > = \left < \phi_s \right > - \left < \phi_s^l \right >$. The size of the largest cluster is a measure of the localization of the $s$-state molecules in the system. For example, in the case where all the $s$-state molecules are assembled together and part of the same inter-connected network, the largest cluster will accommodate all the $s$-state molecules and $ \left < \phi_d  \right > = 0$. Thus, the non-zero value of $\left <\phi_d \right >$ suggests that all the $s$-state molecules are not connected by the neighborhood and hence spatially not localized.      
\begin{figure}[h!]
    \centering
     \begin{subfigure}
         \centering
           \includegraphics[width=\linewidth]{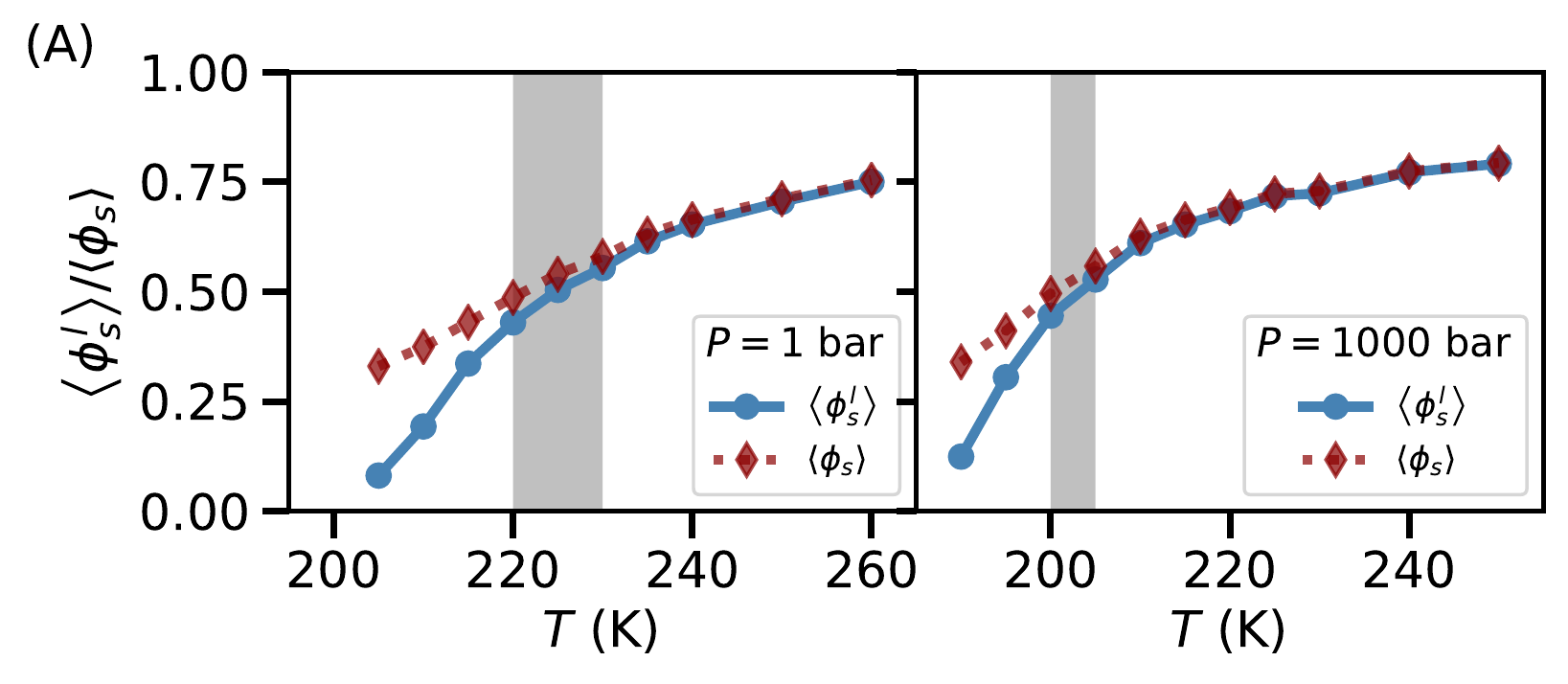}
     \end{subfigure}
     \bigskip
     \begin{subfigure}
         \centering
          \includegraphics[width=\linewidth]{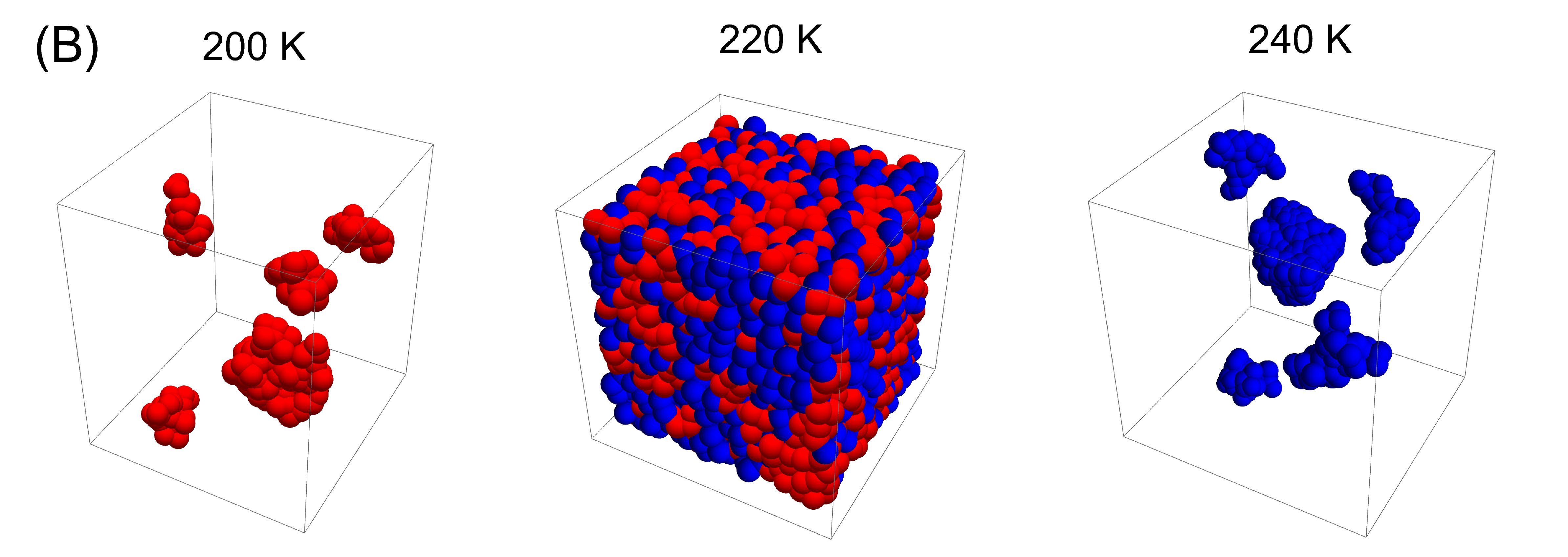}
     \end{subfigure}
    \caption{(A) The $T$-dependent fraction of the $s$-state water molecules, $\left< \phi_s \right>$, along with the fraction of the $s$-state molecules that are part of the largest $s$-cluster, $\left<\phi_s^l\right>$, for $1$ bar and $1000$ bar isobars are shown. The shaded region indicates the ``anomalous region'' defined as the region bounded by the $T_{\kappa_T^{\rm max}}$ on the higher temperature side ($\approx 230$ K and $205$ K for $1$ bar and $1000$ bar isobars, respectively~\cite{anisimov_tip4p}) and the $T_{C_P^{\rm max}}$ on the lower temperature side ($\approx 220$ K and $200$ K for $1$ bar and $1000$ bar isobars, respectively). (B) The representative snapshots showing the temperature (or, composition) dependent spatial distribution of the $s$- and $\varepsilon$-state molecules in a TE configuration at $1$ bar pressure and for $N = 4000$. Red and blue spheres represent the water molecules in the $s$- and $\varepsilon$-states, respectively. We note an interesting crossover from the spatially delocalized $\varepsilon$-clusters to the spatially delocalized $s$-clusters on isobaric cooling below the $T_{C_P^{\rm max}}$.}
    \label{s-cluster}
\end{figure}

Figure~\ref{s-cluster}A shows the $T$-dependent $\left<\phi_s\right>$ and $\left<\phi_s^l\right>$ at pressures $1$ bar and $1000$ bar. The shaded region in the figure indicates the ``anomalous region'' defined as the region bounded by the $T_{\kappa_T^{\rm max}}$ on the higher temperature side ($\approx 230$ K and $205$ K at $1$ bar and $1000$ bar, respectively~\cite{anisimov_tip4p}) and the $T_{C_P^{\rm max}}$ on the lower temperature side ($\approx 220$ K and $200$ K at $1$ bar and $1000$ bar, respectively). As evident from Fig.~\ref{s-cluster}A, almost all the $s$-state molecules are part of the largest $s$-cluster at higher temperatures ($\left < \phi_d \right > \sim 0 $), and the $\varepsilon$-state molecules are delocalized inside the spanning network of the $s$-state molecules (see Fig.~\ref{s-cluster}B). On (isobaric) cooling, both $\left <\phi_s \right >$ and $\left < \phi_s^l \right >$ decrease as a consequence of the conversion of the $s$-state molecules to the $\varepsilon$-state. Interestingly, inside the ``anomalous region'', $T_{C_P^{\rm max}}  < T < T_{\kappa_T^{\rm max}}$, both the $\varepsilon$- and $s$-state molecules form an interpenetrating spanning dynamical network, and on further lowering the temperature ($T <  T_{C_P^{\rm max}}$), the $s$-state molecules get delocalized in the form of smaller clusters inside a relatively rigid spanning network of the $\varepsilon$-state molecules (see Fig.~\ref{s-cluster}B). Thus, the anomalous behavior of the thermodynamic response functions in the supercooled water is a direct consequence of the composition fluctuation and composition-dependent spatial organization of the $s$- (or, $\varepsilon$) state molecules. At this point, it is also worth noting that the observed fragile-to-strong dynamical transition (often referred to as ``dynamical crossover'') in liquid water on cooling below the $T_{C_P^{\rm max}}$~\cite{stanley_pnas_2005, gallo_fsc} could be a consequence of the formation of the relatively rigid spanning network of the $\varepsilon$-state molecules. This observation, however, needs a more careful quantitative validation to unambiguously establish an intimate connection between the anomalous thermodynamic and dynamic behavior of supercooled water.     
%
 \begin{figure}[h]
 \vspace{10pt}
 \begin{center}
       \includegraphics[width=0.9\linewidth]{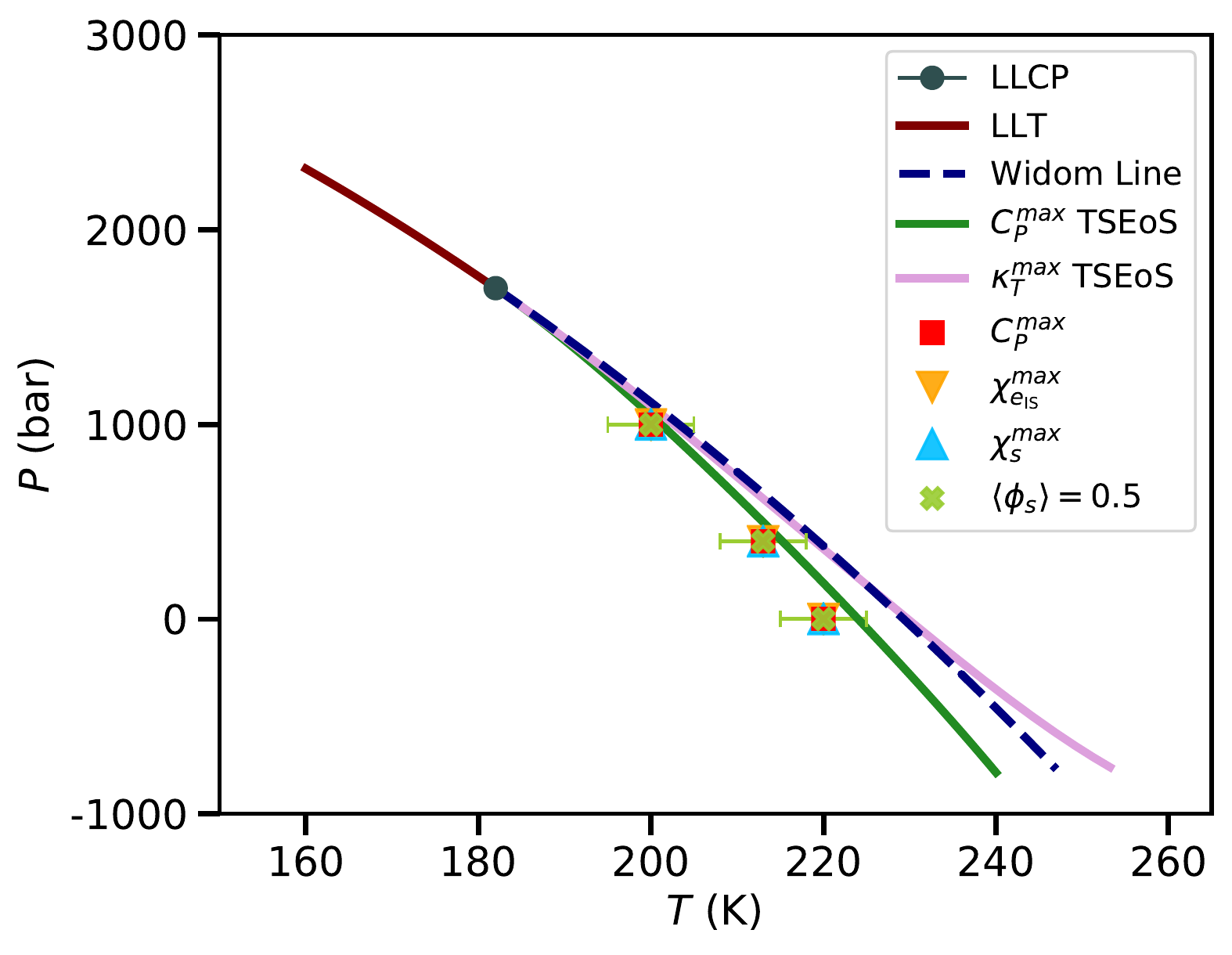}
   \caption{The phase diagram of supercooled TIP4P/2005. The locus of maxima of heat capacity $C_P^{\rm max}$, isothermal compressibility $\kappa_T^{\rm max}$, the Widom line, the LLT line, and the LLCP predicted by the TSEoS~\cite{singh_tip4p_2017} along with the $C_P^{\rm max}$, maximum of the $s$-state population fluctuation $\chi_s^{\rm max}$, the maximum of the IS energy per molecule fluctuation $\chi_{e_{\rm IS}}^{\rm max}$, and the $1:1$ population ratio of the $s$-state and the $\varepsilon$-state molecules ($\left \langle \phi_s \right \rangle = 0.5$) calculated using the TE configurations for isobars $1$ bar, $400$ bar, and $1000$ bar are shown. We note that the $\chi_s^{\rm max}$, $\chi_{e_{\rm IS}}^{\rm max}$, and $1:1$ population ratio of the $s$-state and the $\varepsilon$-state ($\left \langle \phi_s \right \rangle = 0.5$) fall in the close vicinity of the $C_P^{\rm max}$ line predicted by the TSEoS.} 
   \label{phase_diagram}
   \end{center}
\end{figure}
\subsection{\label{sec:7}Phase behavior of TIP4P/2005 water}
In Fig.~\ref{phase_diagram}, we present a phase diagram summarizing the thermodynamic behavior of supercooled TIP4P/2005 water in the $P-T$ plane predicted by the phenomenological TSEoS~\cite{singh_tip4p_2017}, along with the heat capacity maximum $C_P^{\rm max}$, maximum of the $s$-state population fluctuation $\chi_s^{\rm max}$, the maximum of the IS energy per molecule fluctuation $\chi_{e_{\rm IS}}^{\rm max}$, and the $1:1$ population ratio of the $s$-state and the $\varepsilon$-state water molecules ($\left \langle \phi_s \right \rangle = 0.5$) calculated in this work using the TE configurations at pressures $1$ bar, $400$ bar, and $1000$ bar. The $C_P^{\rm max}$, $\chi_s^{\rm max}$, $\chi_{e_{\rm IS}}^{\rm max}$, and $\left \langle \phi_s \right \rangle = 0.5$ points fall in the close vicinity of the $C_P^{\rm max}$ predicted by the TSEoS for all the three isobars studied in this work (Fig.~\ref{phase_diagram}). Thus, this phase diagram also summarizes the intricate relationship between the (anomalous) temperature dependence of $C_P$, the $s$-state (or, $\varepsilon$-state) population fluctuation in the TE configurations, and the $\chi_{e_{\rm IS}}^{\rm max}$. The $\chi_s^{\rm max}$, $\chi_{e_{\rm IS}}^{\rm max}$ and $\left \langle \phi_s \right \rangle = 0.5$ can also be used as alternate markers for the heat capacity anomaly line or the Widom line in the phase plane (see also Fig.~\ref{defect_diagram}B and Fig.~\ref{figure2}).   
\section{Conclusions}\label{conclusions}
In this study, we have employed a predefined structural order parameter free approach to unambiguously identify the two types of interconvertible local states --- entropically-favored $s$-state (or, HDL-like) and energetically-favored $\varepsilon$-state (or, LDL-like) --- with significantly different structural and energetic features in liquid water. This identification is based on the heterogeneous relaxation of the system in the PEL during the steepest-descent energy minimization. The $s$-state molecules undergo more local structural changes compared to the $\varepsilon$-state molecules during the energy minimization. This heterogeneous relaxation in the PEL is characterized by using order parameters inspired by the Edwards-Anderson order parameter~\cite{ea} for the spin-glass transition in frustrated magnetic systems. The $s$- and $\varepsilon$-states share only partial resemblance with the two types of local states, HDL- and LDL-like, respectively, assigned on the basis of the LSI~\cite{lars_2011} and $\zeta$~\cite{tanaka_natcomm} order parameters. We have also shown a $1:1$ composition of the two states at the $T_{C_P^{\rm max}}$ (or, the Widom line). We have further established a direct relationship between the $s$-state population fluctuation $\chi_s$, and the heat capacity $C_P$ anomaly in the supercooled state. We have also studied the composition-dependent spatial distribution of the $s$-state water molecules in the system by employing clustering analysis, and found an interesting crossover from a spanning network-like single $s$-cluster to the spatially delocalized $s$-clusters in the close vicinity of the Widom line on isobaric cooling.  

Although this study does not directly invoke the existence of the LLCP, the results presented here are fully consistent with the thermodynamic interpretation of the supercooled water's anomalies based on the existence of the LLCP and the associated Widom line, which is unambiguously established for the model system studied here (TIP4P/2005)~\cite{pablo_science}. This study also provides alternate markers (in addition to the locus of maxima of thermodynamic response functions) for the Widom line, and unravels the signatures of the anomalous temperature dependence of $C_P$ encoded in the PEL of the system. Thus, this study establishes a direct relationship between the topography of the underlying PEL, the nature of the locally-favored structures in the TE configurations, and the anomalous thermodynamic behavior of the system.  

The prevailing approaches to understand the microscopic origin of bulk phase behavior often involve analysis of either the TE configurations (structures in the free energy landscape) or the ISs (structures in the PEL)~\cite{sastry_nat, stillinger_nat}. The findings in this work bridge the gap between these two approaches by suggesting that the microscopic relaxation mechanism of the TE configurations in the PEL can provide deeper insights into the nature of locally-favored structures and its relationship with the bulk phase behavior. It is also worth noting that the approach discussed here does not provide a parameter-free metric to identify the two types of local states, but it depends on a cut-off radius to define the local environment around a central molecule. This parameter can be interpreted as the length scale associated with the local structural excitations in the PEL, and needs to be carefully chosen to capture the non-trivial relationship between the local structural fluctuations and bulk phase behavior. This parameter also brings in additional generality to the approach (unlike the approaches based on structural order parameters which are mathematically formulated to capture predefined local structures) which can be applied to different model systems, not necessarily molecular, to provide the precise microscopic structural origin of bulk liquid properties.    

\begin{acknowledgments}
R.S.S. thanks Dr. Mantu Santra for helpful discussions. R.S.S. gratefully acknowledges financial support from DST-SERB (Grant No. SRG/2020/001415) and Indian Institute of Science Education and Research (IISER) Tirupati. A.M. and G.R. acknowledge financial support from IISER Tirupati. The computations were performed at the IISER Tirupati computing facility and at PARAM Brahma at IISER Pune.
\end{acknowledgments}

\bibliographystyle{apsrev4-2}
\bibliography{water_is}

\pagebreak
\widetext
\begin{center}
\textbf{\large Supplementary Materials}
\end{center}
\setcounter{equation}{0}
\setcounter{figure}{0}
\setcounter{table}{0}
\setcounter{page}{1}
\setcounter{section}{0}
\makeatletter
\renewcommand{\theequation}{S\arabic{equation}}
\renewcommand{\thefigure}{S\arabic{figure}}

\subsection{\label{subsec:level1} Sensitivity of the $s$- and $\varepsilon$-state populations and their fluctuations on the choice of $\boldsymbol r_{OO}$}
\begin{figure*} [h]
    \centering
     \begin{subfigure}
         \centering
         \includegraphics[scale=0.39]{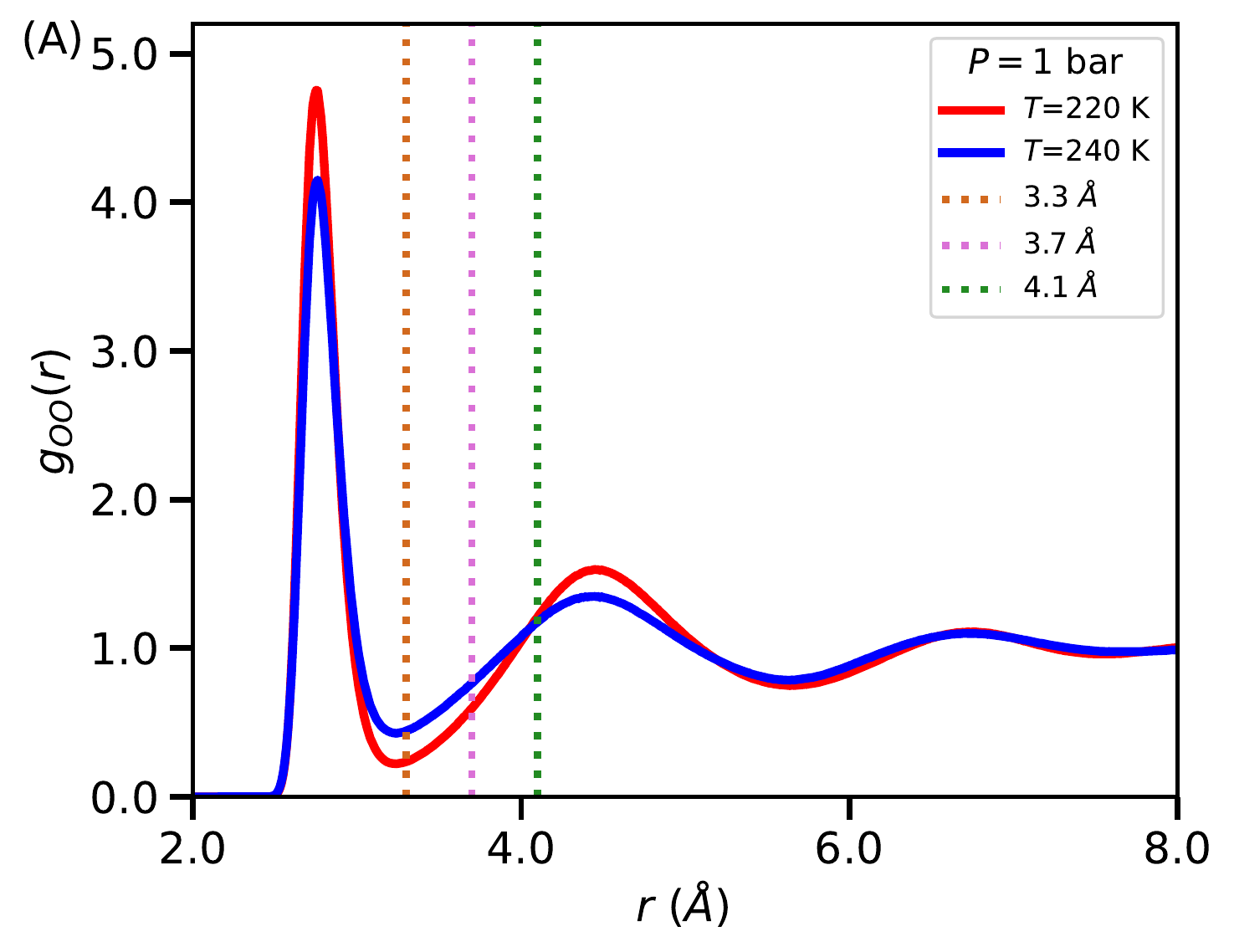}
     \end{subfigure}
     \begin{subfigure}
         \centering
         \includegraphics[scale=0.39]{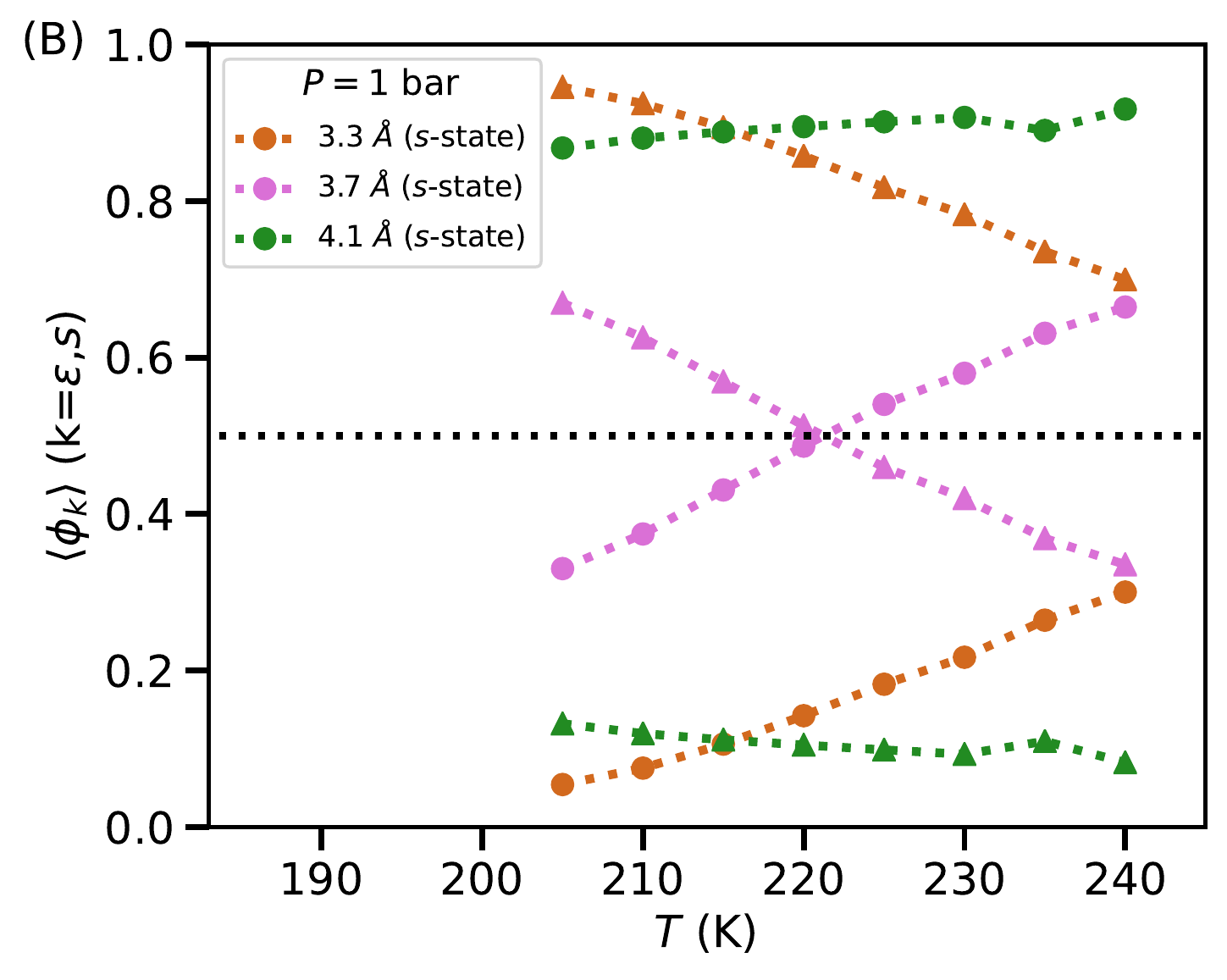}
     \end{subfigure}
     \begin{subfigure}
         \centering
         \includegraphics[scale=0.39]{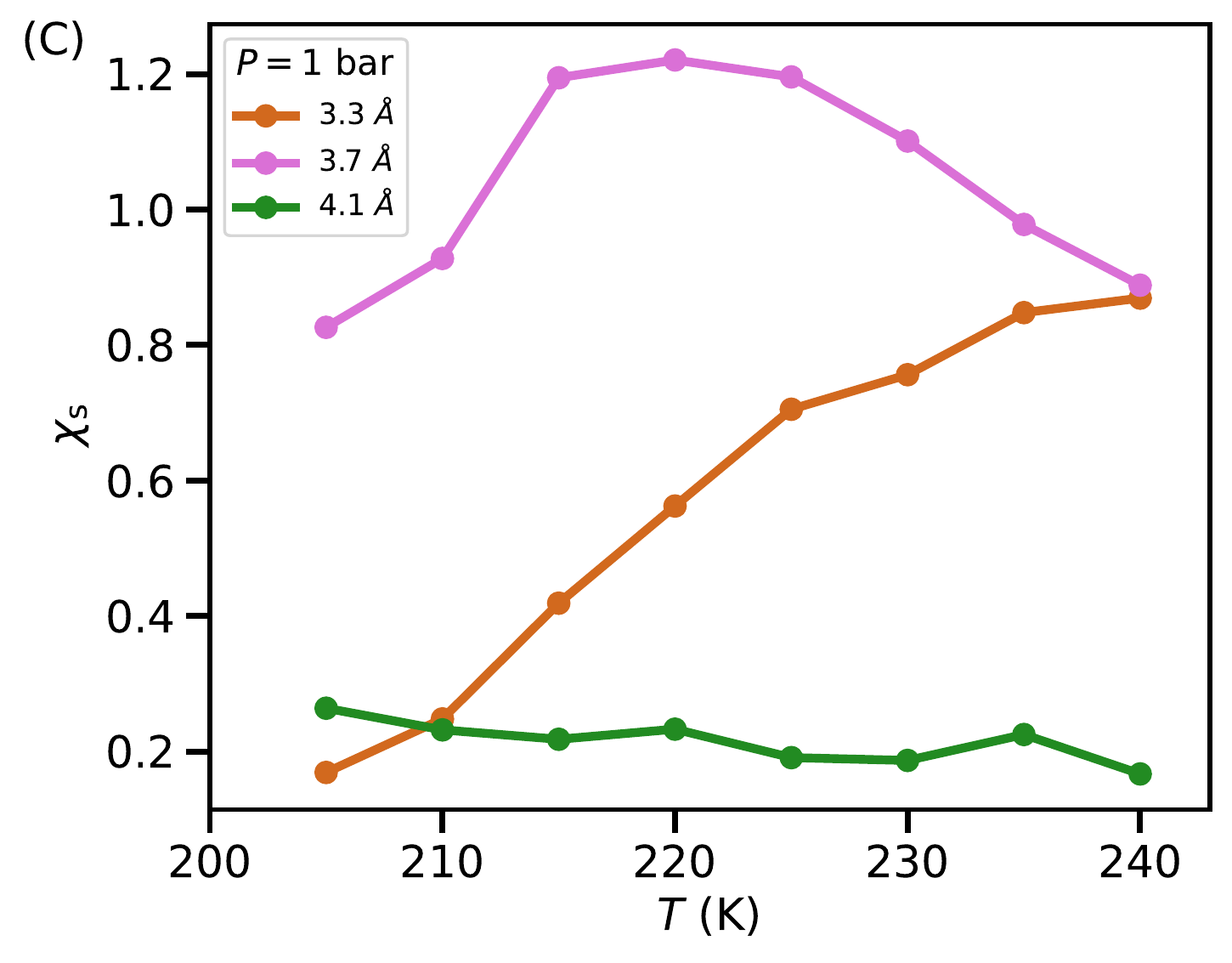}
     \end{subfigure}
    \caption{(A) The oxygen-oxygen pair distribution function ($g_{OO} (r)$) of the TIP4P/2005 water at pressure $1$ bar and at temperatures $220$ K and $240$ K. Three different oxygen-oxygen radial cut-off distances, $r_{OO}$, used to define the local environment (list of neighboring water molecules) -- $3.3~\AA$ (strictly the first shell), $3.7~\AA$ (partly includes the second shell water molecules), and $4.1~\AA$ (closer to the second shell maximum) --  are shown with orange, purple and green dotted lines, respectively. (B) The $T$-dependent average fraction of the two types of local states in the TE configurations ($\left \langle \phi_k \right \rangle$, $k = s$, and $\varepsilon$) for different $r_{OO}$ values at pressure $1$ bar. The dotted black line indicates $\left \langle \phi_s \right \rangle = \left \langle \phi_{\varepsilon} \right \rangle = 0.5$ line. (C) The $T$-dependent $s$-state population fluctuation $\chi_s$ in the TE configurations for different $r_{OO}$ values is shown. We note the non-monotonic temperature dependence of $\chi_s$ for $r_{OO} = 3.7~\AA$, and monotonic for $3.7~\AA$ and $ 4.1~\AA$ cut-off distances.}
    \label{fig1s}
\end{figure*}

To probe the sensitivity of the $s$ (or, $\varepsilon$)-state population and its fluctuation (results presented in Fig.~\ref{defect_diagram} and Fig.~\ref{figure2}A of the main text) on the choice of the oxygen-oxygen cut-off distance, $r_{OO}$, used to define the local environment (list of neighboring water molecules) around a central water molecule, we calculated the $T$-dependent population of the $s$- and $\varepsilon$-state molecules ($\left \langle \phi_k \right \rangle$, $k = s$, and $\varepsilon$) for three different cut-off values, $r_{OO} = 3.3~\AA$ , $3.7~\AA$ and $4.1~\AA$. The $r_{OO} = 3.3~\AA$ includes strictly the first shell water molecules. The cut-off distances $r_{OO} = 3.7~\AA$ and $4.1~\AA$ also include water molecules from the second shell, and hence, enable us to probe the local structural changes involving partly the second shell molecules (see Fig.~\ref{fig1s}A) during the steepest-descent energy minimization. The two ($s$ and $\varepsilon$) states in liquid water are identified following the protocol outlined in Section~\ref{sec:1} of the main text. The Figs.~\ref{fig1s}B and~\ref{fig1s}C show the sensitivity of the $T$-dependent $s$-state population (along with the $\varepsilon$-state population, $\left \langle \phi_{\varepsilon} \right \rangle = 1 - \left \langle \phi_s \right \rangle$) and its fluctuation, $\chi_s$ to the choice of $r_{OO}$ value at $1$ bar pressure in the TE configurations. We note that the $1:1$ fraction of the $s$- and $\varepsilon$-state molecules are obtained near the $T_{C_P^{\rm max}}$ ($220$ K) only for $r_{OO} = 3.7~\AA$. Also, $\chi_s$ shows a maximum near to the $T_{C_P^{\rm max}}$ only for $r_{OO} = 3.7~\AA$. For $r_{OO} = 3.3~\AA$, $\chi_s$ decreases monotonically on lowering the temperature, suggesting that the local (tetrahedral) geometry involving the first shell molecules gradually gets more stable at lower temperatures. For $r_{OO} = 4.1~\AA$,  $\chi_s$ is almost insensitive to the change in temperature.  
\begin{figure*}
    \centering
     \begin{subfigure}
         \centering
         \includegraphics[scale=0.55]{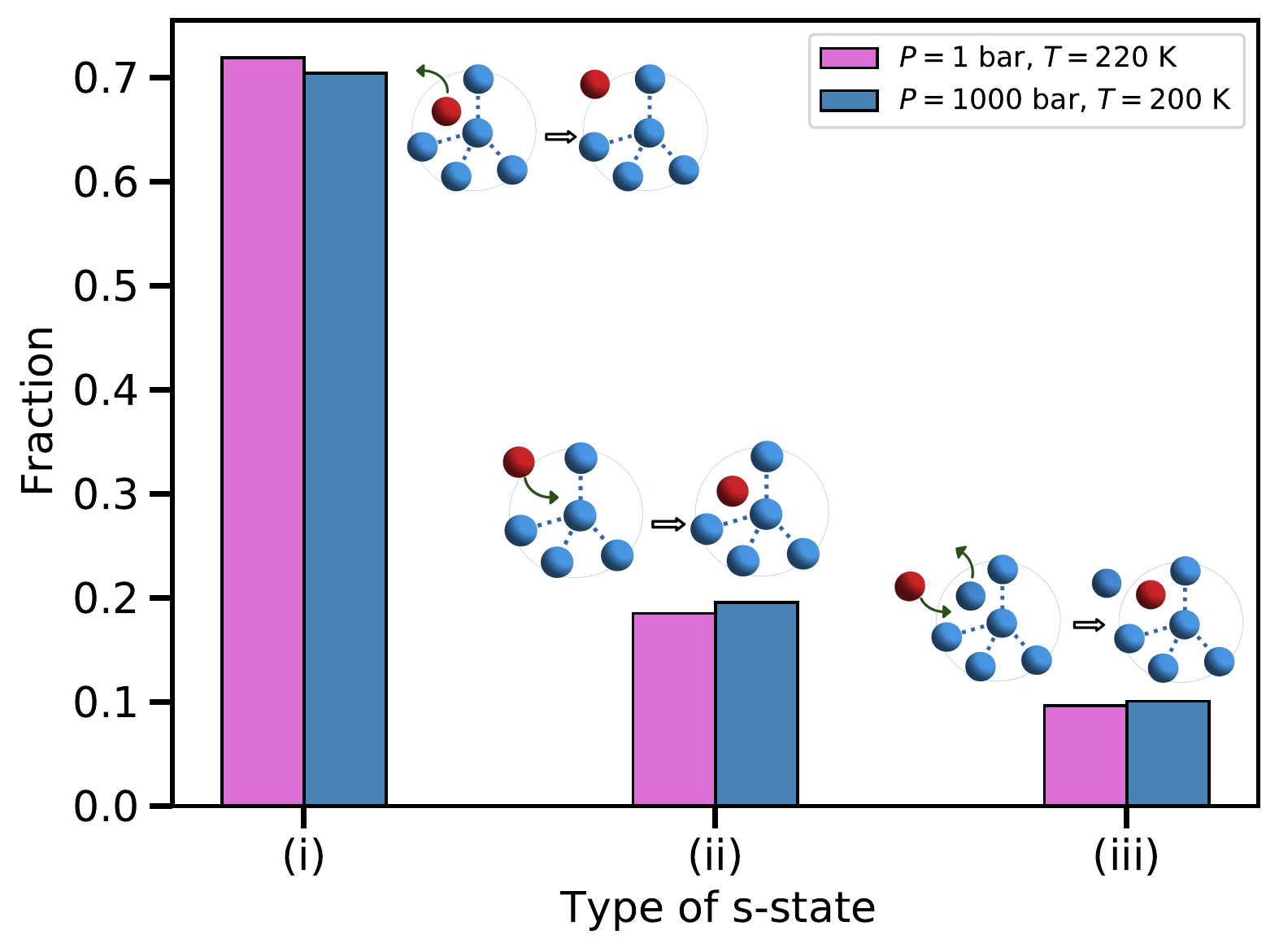}
     \end{subfigure}
    \caption{The fraction of the different types of entropically-favored $s$-state water molecules at pressures $1$ bar and $1000$ bar and at temperatures $220$ K and $200$ K, respectively ($T_{C_P^{\rm max}}$ for $1$ bar and $1000$ bar isobars, respectively)} 
    \label{fig2xs}
\end{figure*}
\begin{figure*}
    \centering
     \begin{subfigure}
         \centering
         \includegraphics[scale=0.6]{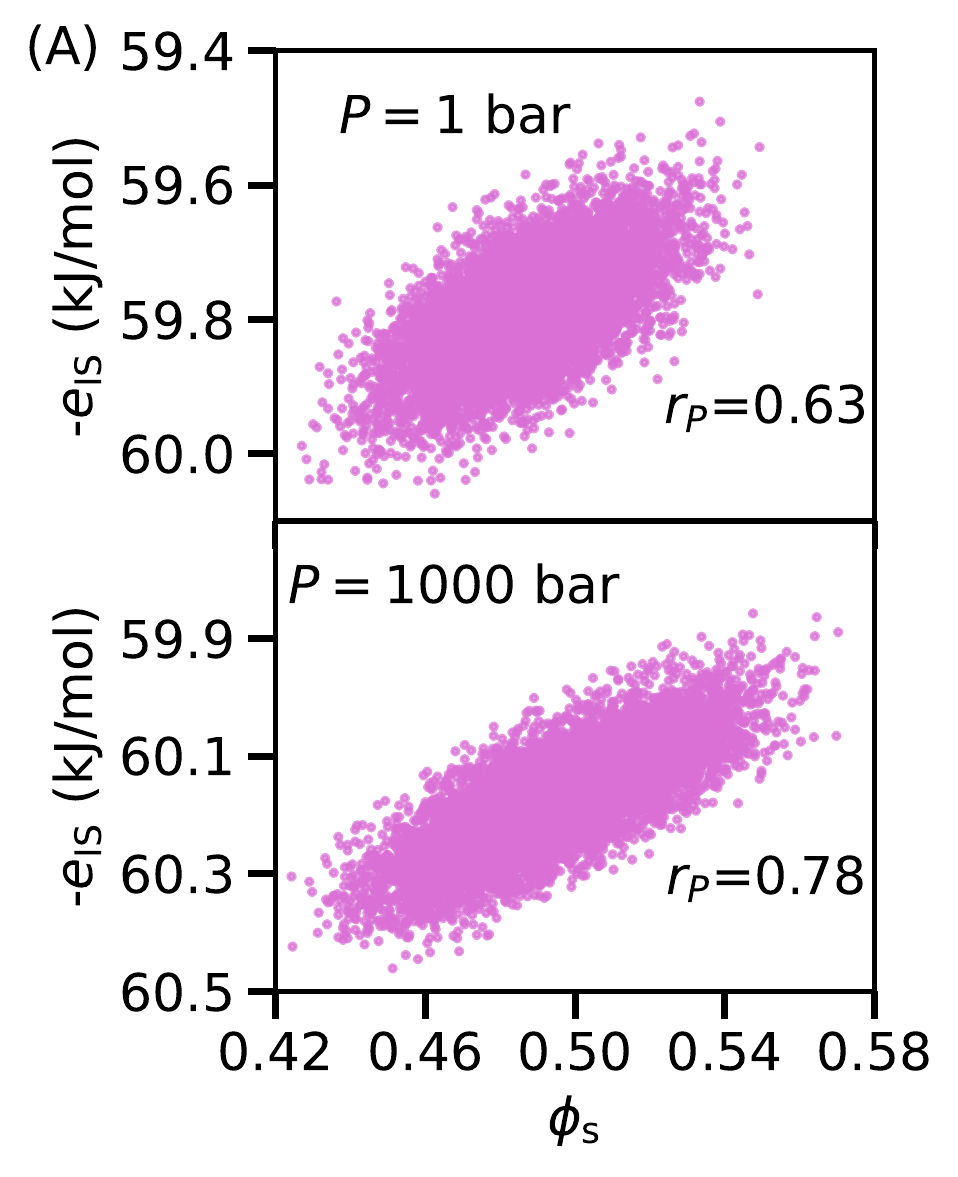}
     \end{subfigure}
     \begin{subfigure}
         \centering
         \includegraphics[scale=0.6]{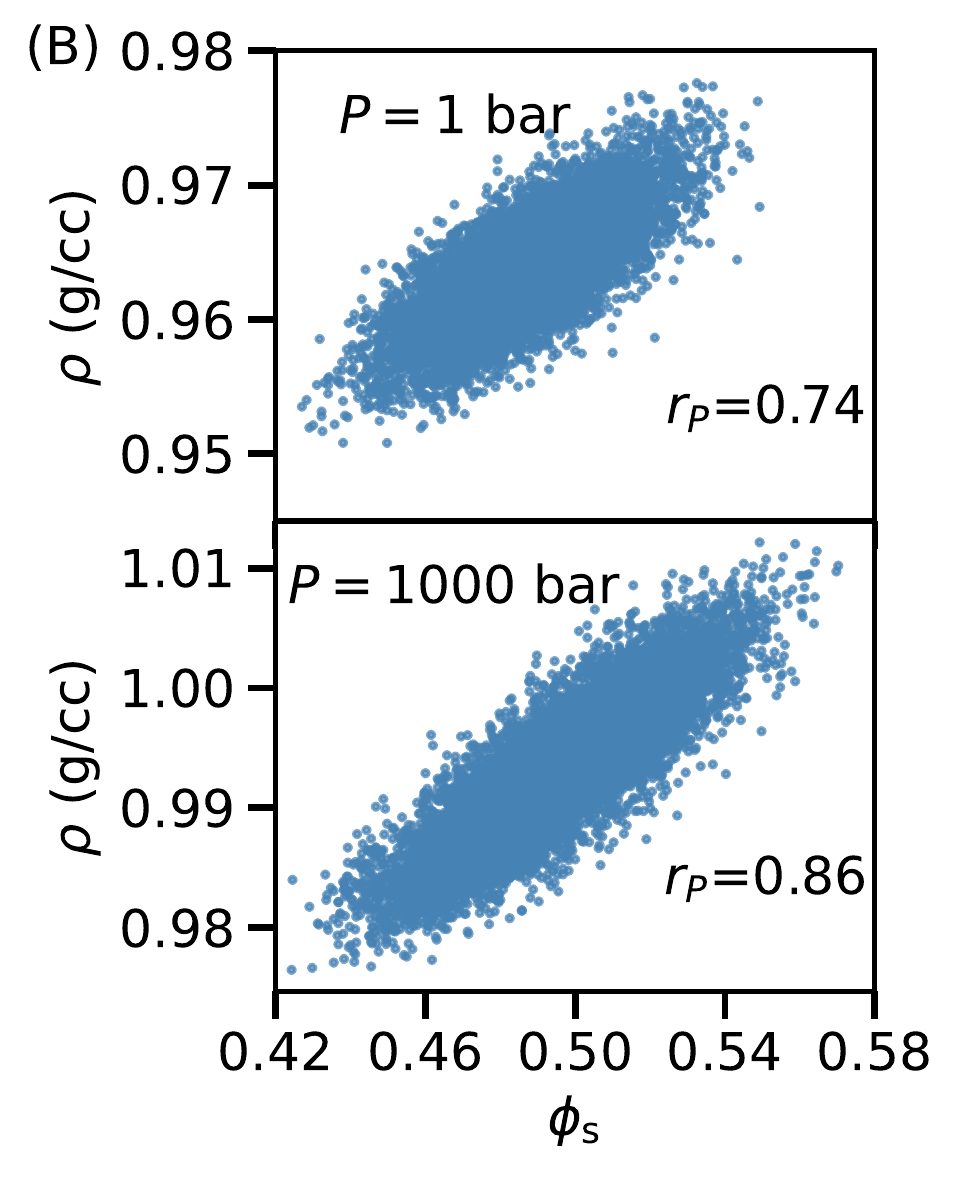}
     \end{subfigure}
    \caption{The correlation between (A) the fraction of the $s$-state molecules in a TE configuration ($\phi_s$) and the (potential) energy per molecule of the corresponding IS ($e_{\rm IS}$), and (B) the mass density ($\rho$) and $\phi_s$ at $1$ bar and $1000$ bar pressures and at temperatures $220$ K and $200$ K, respectively ($T_{C_P^{\rm max}}$ for $1$ bar and $1000$ bar isobars, respectively). $r_p$ indicates the Pearson's correlation coefficient. We note a strong positive correlation between $\phi_s$ and $e_{\rm IS}$ as well as between $\phi_s$ and $\rho$ for both the isobars at their respective $T_{C_P^{\rm max}}$. We also note that the correlation ($r_p$ value) increases on increasing the pressure of the system.}
    \label{fig2s}
\end{figure*}

\subsection{\label{subsec:level2} Distinct types of entropically-favored local ($s$-state) configurations}
In Fig.~\ref{fig2xs}, we show the relative fraction of the different types of entropically-favored $s$-state water molecules at pressures $1$ bar and $1000$ bar and at temperatures $220$ K and $200$ K, respectively ($T_{C_P^{\rm max}}$ for $1$ bar and $1000$ bar isobars, respectively). On potential energy minimization, the majority of the $s$-state molecules (ca. $70 \%$) lose a nearest neighbor (water molecule with $r_{OO} < 3.7~\AA$ from the central molecule) and become tetrahedrally coordinated (referred to as type (i)), ca. $20 \%$ $s$-state molecules gain a neighbor (type (ii)), and ca. $10 \%$ molecules exchange a neighbor with the surrounding (type (iii)).    

\subsection{\label{subsec:level3}Correlation between the inherent structure (IS) energy, the $s$-state population, and the density of the system}
In order to understand the correlation between the $s$-state population in a TE configuration and the (potential) energy of the corresponding IS basin in the PEL, we calculated the Pearson's correlation coefficient ($r_p$) between the potential energy per molecule of the IS, $e_{\rm IS}$, and the fraction of the $s$-state molecules, $\phi_s$ in the corresponding TE configuration. Pearson's correlation is defined in such a way that the value is always between $-1$ and $1$, $-1$ being strong negative correlation and $1$ being strong positive correlation. Fig.~\ref{fig2s}A shows a strong positive correlation between $e_{\rm IS}$ and $\phi_s$ at $1$ bar and $1000$ bar pressures and at temperatures $220$ K and $200$ K, respectively ($T_{C_P^{\rm max}}$ for $1$ bar and $1000$ bar isobars, respectively). As the pressure increases, the correlation (the value of $r_p$) also increases. In Fig.~\ref{fig2s}B, we show the correlation between the mass density $\rho$ and $\phi_s$ at the conditions reported above. We again note a strong positive correlation for both the isobars at their respective $T_{C_P^{\rm max}}$. 

\subsection{\label{subsec:level4}Local free energy surface}
\begin{figure*}
    \centering
     \begin{subfigure}
         \centering
         \includegraphics[scale=0.5]{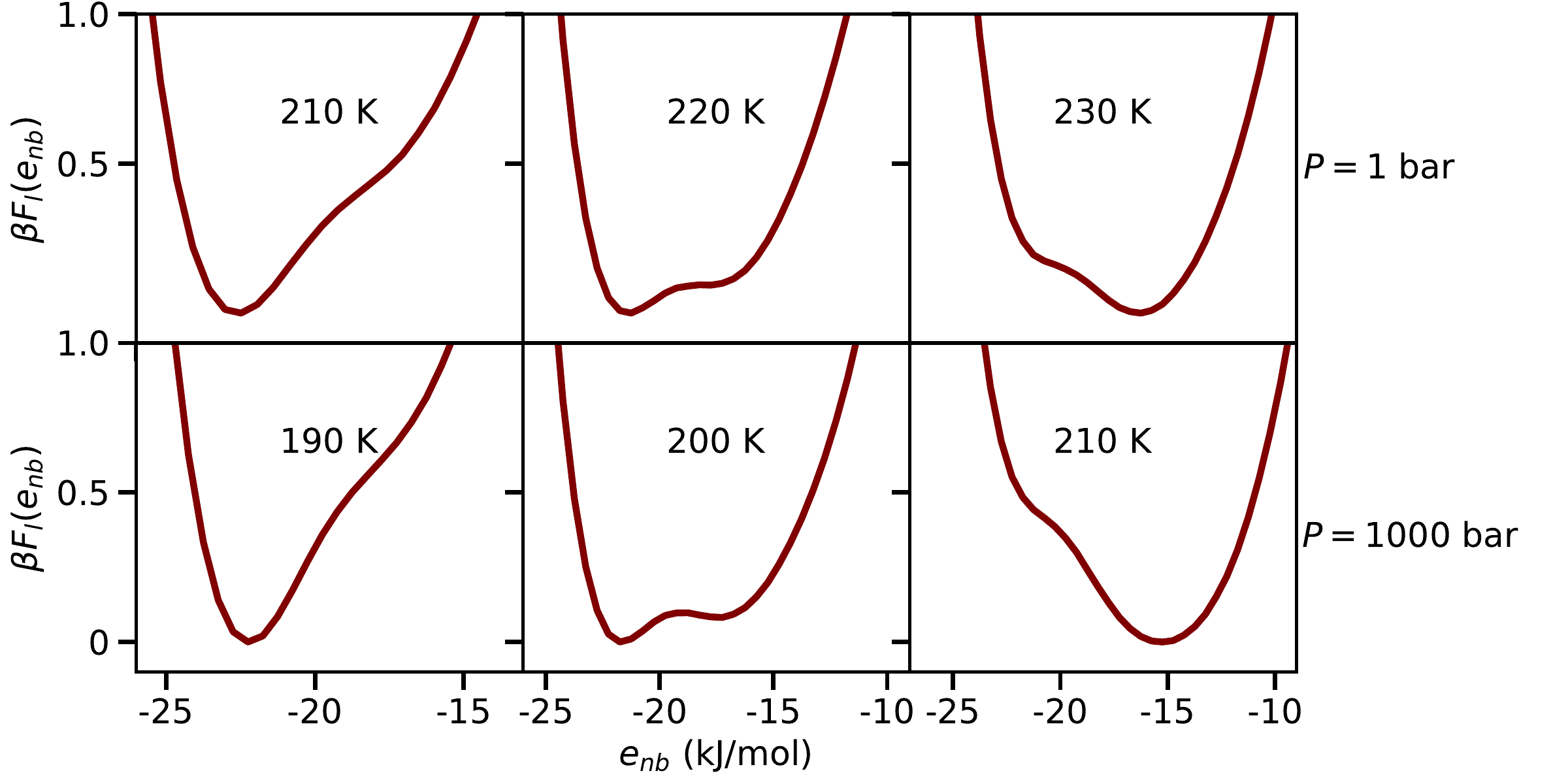}
     \end{subfigure}
    \caption{The $T$-dependent scaled local free energy surface ($\beta F_l (e_{nb})$) for $1$ bar and $1000$ bar isobars and at temperatures in the close vicinity of their respective $T_{C_P^{\rm max}}$, $220$ K and $200$ K. The scaled local free energy is defined as, $\beta F_l(e_{nb}) = -\ln \left[P(e_{nb})\right]$, $\beta = 1/k_{\rm B}T$. We note a two-well feature of $F_l(e_{nb})$ at the $T_{C_P^{\rm max}}$.} 
    \label{fig3s}
\end{figure*}

The bimodal distribution of the potential energy per neighbor, $e_{nb}$ (see Section~\ref{sec:4} of the main text) in the close vicinity of the $T_{C_P^{\rm max}}$ suggests a two-well nature of the underlying local free energy surface, defined as, $F_l(e_{nb}) = -k_{\rm B}T \ln \left[P(e_{nb})\right]$, where $P(e_{nb})$ is the local interaction energy per neighbor distribution function. In Fig.~\ref{fig3s}, we show the $T$-dependent evolution of the scaled local free energy surface, $\beta F_l(e_{nb})$ ($\beta = 1/k_{\rm B}T$) at pressures $1$ bar and $1000$ bar. As expected, $\beta F_l(e_{nb})$ shows a two-well feature at the $T_{C_P^{\rm max}}$ (or, near the Widom line). This observation suggests that the $s$-state and the $\varepsilon$-state molecules share distinct energetic features and $e_{nb}$ can also be used as an order parameter to identify the two types of local states in liquid water.  
\begin{figure*}
    \centering
     \begin{subfigure}
         \centering
         \includegraphics[scale=0.7]{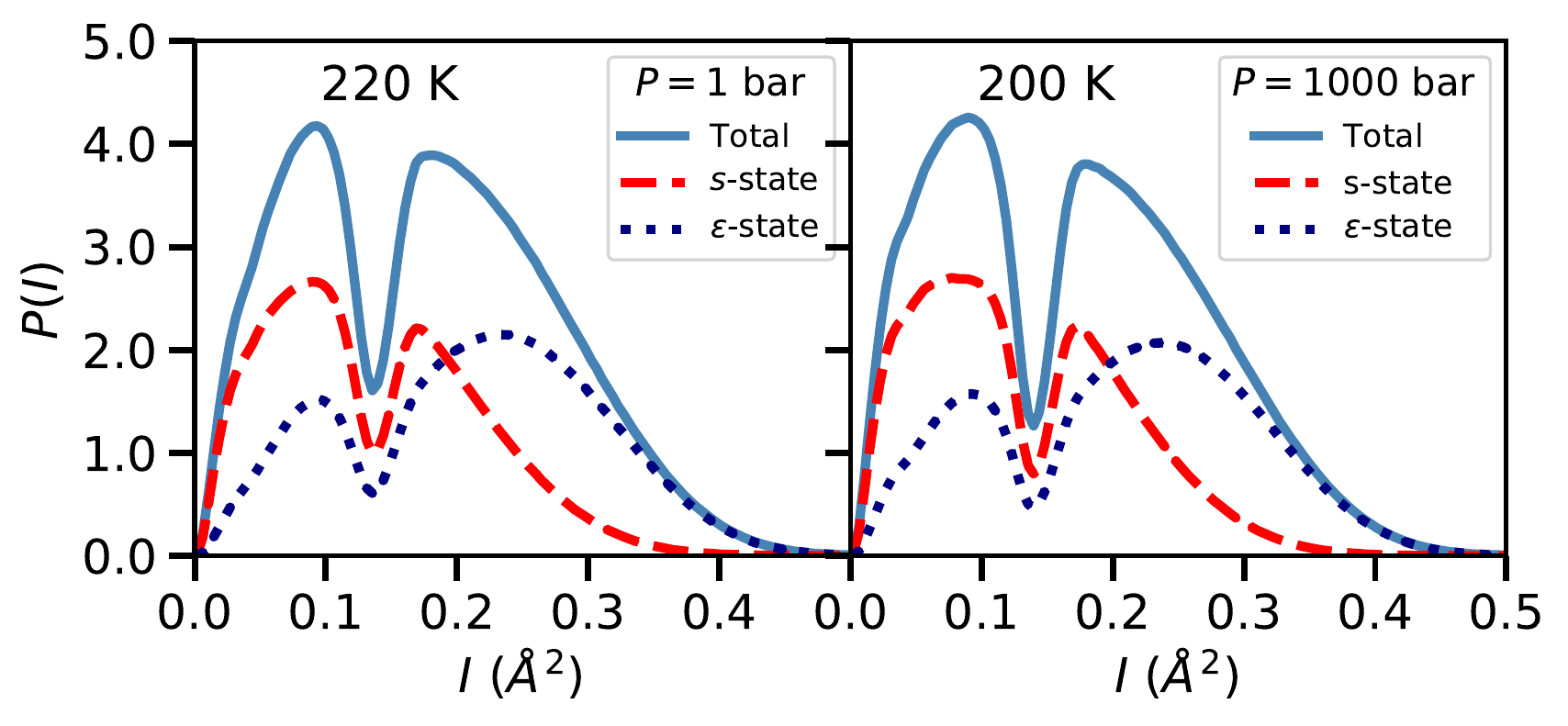}
     \end{subfigure}
    \caption{The LSI distribution for the $s$-state (red dashed line), the $\varepsilon$-state (blue dotted line), and for all the water molecules (solid blue line) in the ISs at $1$ bar and $1000$ bar pressures and at temperatures $220$ K and $200$ K, respectively ($T_{C_P^{\rm max}}$ for $1$ bar and $1000$ bar isobars, respectively). We note a bimodal distribution for both the states, suggesting that the $s$- and $\varepsilon$-state molecules share only partial resemblance with the LDL- and HDL-like molecules assigned on the basis of the LSI in the PEL of the system.}
    \label{fig5s}
\end{figure*}
\begin{figure*}
    \centering
     \begin{subfigure}
         \centering
         \includegraphics[scale=0.7]{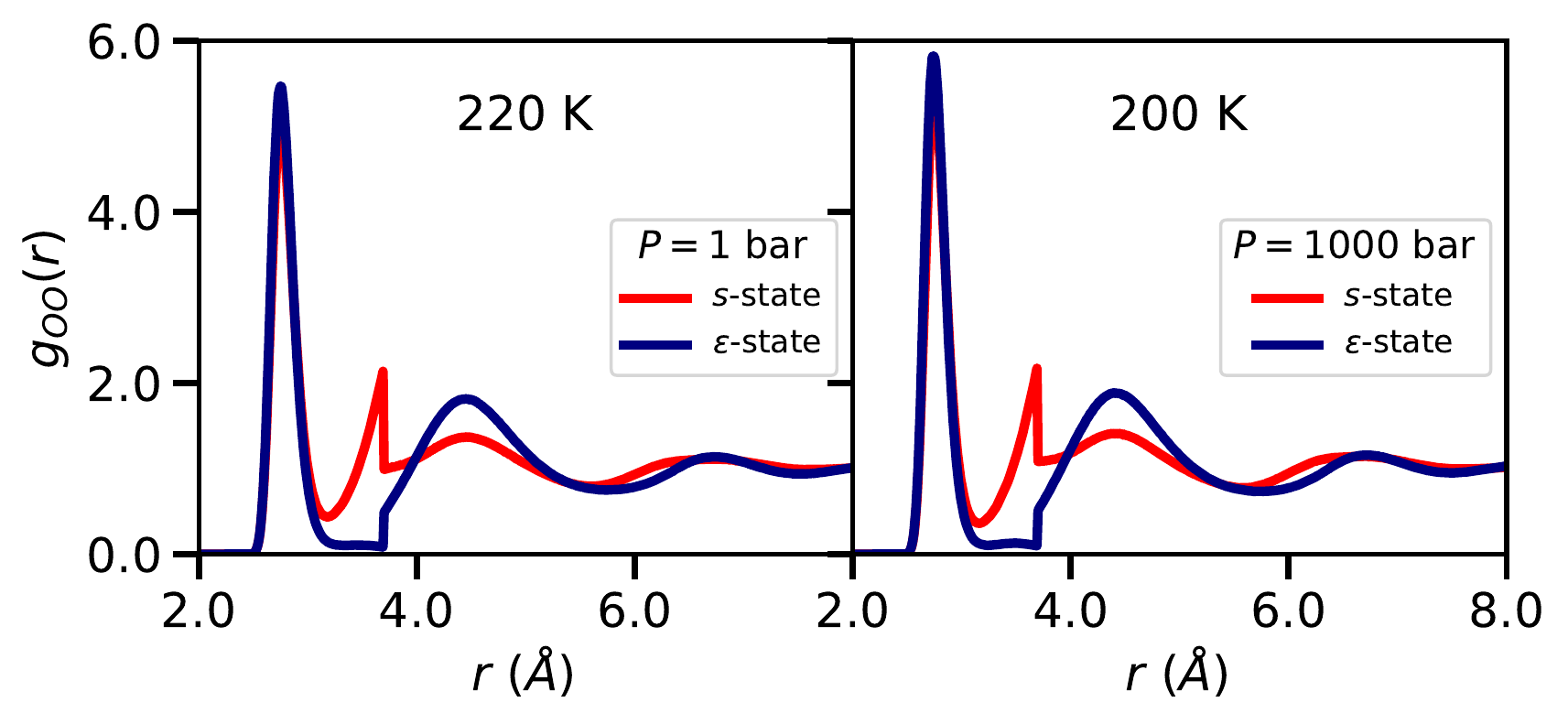}
     \end{subfigure}
    \caption{The two-body oxygen-oxygen correlation function ($g_{OO}$) for the $s$-state and the $\varepsilon$-state water molecules at $1$ bar and $1000$ bar pressures and at temperatures $220$ K and $200$ K, respectively ($T_{C_P^{\rm max}}$ for $1$ bar and $1000$ bar isobars, respectively). We note a near-complete separation of the first and second coordination shells for the case of $\varepsilon$-state molecules, and the presence of interstitial water molecules in between the first and second shells for the case of $s$-state molecules.}
    \label{fig4s}
\end{figure*} 

\subsection{\label{subsec:level6} LSI distribution for the $s$- and the $\varepsilon$-state water molecules in the IS}
The two types of local states in liquid water have often been assigned using the LSI order parameter which shows a pronounced bimodality for the ISs (see, for example, Refs.~\cite{lsi_spce, lars_2011}). The bimodal LSI distribution gives us a unique order parameter cut-off value to unambiguously identify the two types of local states in the ISs (or, in the PEL of the system). In Fig.~\ref{fig5s}, we show the LSI distribution for the $s$-state, the $\varepsilon$-state, and for all the molecules in the ISs to check whether the LDL- and HDL-like local states assigned on the basis of the LSI share the same structural features as the $\varepsilon$- and $s$-states, respectively, in the PEL. We note, however, a bimodal LSI distribution for both the states at the $T_{C_P^{\rm max}}$, suggesting that the $\varepsilon$- and $s$-state molecules share only partial resemblance with the LDL- and HDL-like molecules, respectively, assigned on the basis of the LSI in the PEL of the system.  

\subsection{\label{subsec:level5}Two-body correlation function for the $s$- and $\varepsilon$-state water molecules}
In Fig.~\ref{fig4s}, we show the two-body oxygen-oxygen correlation function ($g_{OO}$) for the $s$- and $\varepsilon$-state molecules at $1$ bar and $1000$ bar pressures and at temperatures $220$ K and $200$ K, respectively ($T_{C_P^{\rm max}}$ for $1$ bar and $1000$ bar isobars, respectively). We note a near-complete separation of the first and second shells for the case of $\varepsilon$-state molecules, and the presence of interstitial water molecules in between the first and second shells for the case of $s$-state molecules. This suggests that the networks of the $s$-state and the $\varepsilon$-state water molecules show distinct (local) structural features. It is also remarkable to note that these distinct local structural features are captured here without using any predefined structural order parameter that is mathematically formulated to capture these two types of structural features (such as, LSI~\cite{lsi} or $\zeta$ index~\cite{ tanaka_natcomm}).

\end{document}